\documentclass{cernrep}
\usepackage{a4}
\usepackage[bookmarks, colorlinks=true, linktoc=page, pdftex, linkcolor=black, citecolor=black, urlcolor=blue]{hyperref}

\usepackage{fancyhdr}
\fancyhfoffset{4 mm}
\fancypagestyle{ARTTITLE}{%
\fancyhf{} 
\lhead{\small{Proceedings of the CERN--Accelerator--School course: \it{Introduction to Accelerator Physics}}}
\lfoot{Available online at \url{https://cas.web.cern.ch/previous-schools}}
\rfoot{\thepage\hspace*{3mm}}

}
\begin{document}

\title{Machine and People Protection}
\author{Peter Forck}
\institute{GSI Helmholtz-Zentrum f\"ur Schwerionenforschung, Darmstadt, Germany}

\begin{abstract}
{This contribution acts as an introduction to the requirements of the machine protection system. As the first step, the basics interactions of fast charged particles, neutrons and $\gamma$-rays with matter are summarized. The architecture of a machine protection system based on beam loss detection is described. Personal safety issues and personal dosimetry are discussed, including the concept of radiation shielding.}
\end{abstract}



\maketitle
\thispagestyle{ARTTITLE}

\section{Introduction}
Any technical installation must respect the general principle of safety for three reasons:
\begin{itemize}
\item Any person shall be protected against negative influence on physical and psychological health.
\item The environment shall be protected from being unnecessarily loaded and potentially destroyed.
\item The accelerator shall be protected against destruction to enable the facility's safe operation.
\end{itemize}
Legal acts regulate the first two topics, and it should be in the interest of a society to achieve the best possible settings to protect human life and environmental diversity. For an accelerator facility, the risk is related to the destruction of components, either directly by the beam or by other means of failure. People must be protected against direct or indirect negative impact, and the environment must be shielded against unnecessary pollution, including the release of ionizing radiation and activation. The associated laws and associated ordinances should provide an appropriate balance between people and environmental protection and benefits of scientific, technical and economic achievements. Generally, the goal of a~machine protection system is to:
\begin{itemize}
	\item quantize possible device and beam quality degradation
	\item broadcast a general warning to the relevant system of the accelerator facility
	\item trigger possible counteractions or eventually dump the beam and prevent further beam production
	\item enable a clear statement on the reason for a failure by so-called post-mortem analysis.
\end{itemize}
The personal safety system's goal is to protect the employees, users, and general public from being confronted by any hazard in case of an accelerator against above-threshold exposure to ionizing radiation.

\begin{figure}
   	\centering \includegraphics*[width=120mm,angle=0]
   	{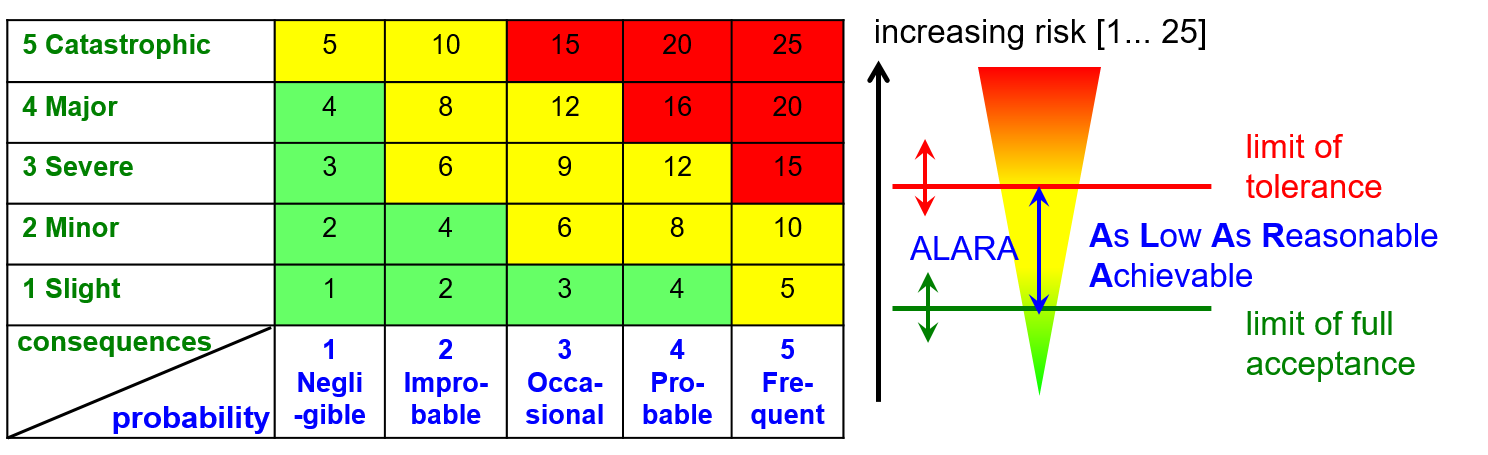} \vspace*{-3mm}
   	\caption{Risk assessment matrix related to the probability of occurrence (horizontal axis) and consequences (vertical axis). Limits of acceptance (green, matrix value 4) and tolerances (red, matrix value 12) depend on the application.}
   	\label{risk_alara}
\end{figure}

Damages of accelerator components or activation of equipment are potential risks for an accelerator facility's operation. The phrase risk is not a physical quantity but frequently used in facility management and can be defined as the product of unintended event occurrence probability and the related consequences as
\begin{equation}\label{key}
	\mbox{Risk = Probability x Consequences}~~.
\end{equation}
A related matrix is depicted in Fig.~\ref{risk_alara}. For the general understanding of the phrase risk, the probability of occurrence is described either in words or quantitatively fixed. The description of consequences depends significantly on the accelerator goal. Commonly two levels are defined: The first is related to the general acceptance of a risk as an event either occurs seldom or has no significant consequences; this defines the 'green area'. The other boundary is a limit of tolerance, where the accelerator operation must be stopped, and an extended shutdown time might be the consequence. The assessment of both limits depends significantly on the accelerator's objective: For accelerators of medical purposes, e.g. patient treatments, the limit of tolerance is stringent, as failures might hurt the patient directly. Accelerators used for basic research allow for much higher risk as novel techniques might be applied. However, the protection of employees and users must be guaranteed as foreseen by legal acts.  Independent of its application, any accelerator design and operation task is related to minimizing risks, i.e. making it As Low As Reasonably Achievable, the ALARA principle. The consequences of a failure are more challenging to assess, as quite different measures can be applied. One possibility is a measure of cost and personal efforts in terms of a damaged component replacement. Other measures could be the time lag of scientific and technical knowledge related to basic research or, generally, a facility's reputation. For accelerators within a production chain, e.g. for radioactive material or patient treatment, further consequences must be considered \cite{Nordt-CAS2017,Willeke-CAS2016,Biarrotte-CAS2013}.

\begin{figure}
	\vspace*{-3mm} \centering \hspace*{-3mm}  \includegraphics*[width=165mm,angle=0]
	{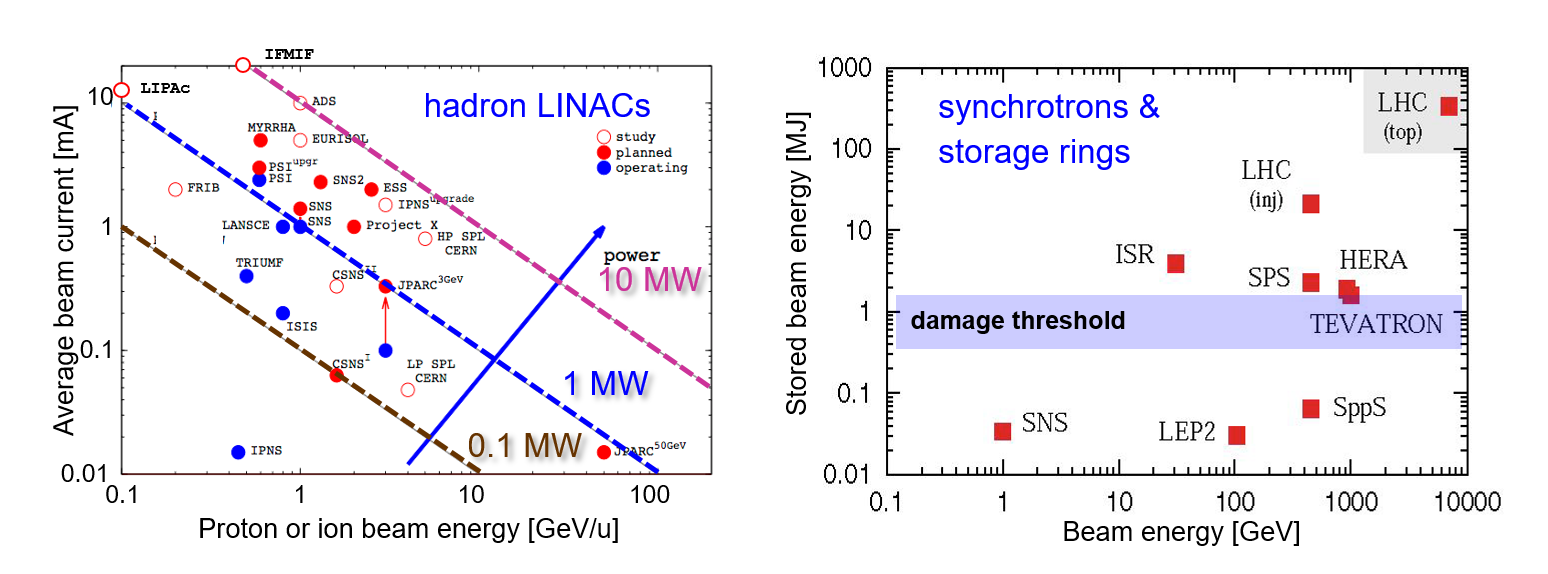}   \vspace*{-7mm}
	\caption{Left: Chart of average beam current and beam energy for various proton and ion LINACs with an indication of the average beam power. Right: Stored beam energy as a function of beam energy for various synchrotrons \cite{Schmidt-CAS2014}.}
	\label{beampower-all-acc}
\end{figure}

For LINACs, the beam energy is gradually increasing after each cavity; the total power $P$ carried by the beam is composed of the beam energy $E_{kin}$, the number of accelerated particle $N$ and the pulse length of beam production $t$ as $P=N \cdot E_{kin}/t$ or $P= I \cdot E_{kin} /e$ expressed in beam current $I$ and elementary charge $e$. To compare different LINACs, the time-averaged beam current is frequently used. Figure~\ref{beampower-all-acc} depicts the average beam power for various proton and ion LINACs. It shows that present-day or near-future proton facilities almost reach 10 MW. For circular accelerators, the stored beam's energy reaches values up to 300 MJ for LHC final energy. For other proton synchrotrons, the stored energy reaches almost 10 MJ. If such beams would be kicked out of the synchrotron within one revolution and dumped inside a metal block in a focused manner, this value is above any damage threshold depicted as a range in Fig.~\ref{beampower-all-acc}. For damage prevention, machine protection systems (MPS) are installed in any high power accelerator to stop beam production or initialize a safe beam dump in case of any malfunction.

In the second chapter, we will summarize the basis of particle interaction with matter. We distinguish between atomic interactions, i.e. the Coulomb interaction between projectiles and target electrons and nuclear reactions related to the strong interaction. The atomic interaction leads to the emission of charged particles and photons; most of these secondary particles are absorbed in the stopping material leading to a temperature increase of surrounding material. With intense beams, the threshold for material destruction can easily be reached, or, for super-conducting components, a quench can be initialized.  The nuclear interaction can lead to radioactive nuclei and activation, followed by emission of $\alpha$, $\beta$ or $\gamma$ ionizing radiation, which restricts human access to the components. Any radiation might lead to modifications of the material properties and a related change in their functionality which counts as a risk for the accelerator operation.

The third chapter concerns the general design of a machine protection system and its technical realization. In almost all cases, the machine protection is based on beam loss monitors; various kinds are used depending on beam species, energy, beam current and required sensitivity.

The fourth chapter summarizes the active and passive methods for people protection. The fundamental radiological quantities are presented. The threshold doses of different radiation zones and their monitoring are described. The legally allowed levels are compared to natural doses. The basic properties concerning the accelerator shielding are introduced.

This text is based on excellent descriptions concerning machine and people protection \cite{Schmidt-CAS2014, Forkel-CAS2013} as well a closely related topics, like risk assessment  \cite{Nordt-CAS2017,Willeke-CAS2016,Biarrotte-CAS2013}, particle-matter interaction \cite{Lechner-CAS2017, Mokhov-CAS2016}, material modification \cite{Kiselev-CAS2013,bertarelli-CAS2016} and active collimation systems \cite{redaelli-CAS2016, Wenn-CAS2016}. Further information on various aspects of machine protection is available from the year 2014 'Joint International Accelerator School on Beam Loss and Accelerator Protection' \cite{CAS2016}.

\section{Beam material interactions with relevance for accelerators}
\label{section-beam-inter}
\subsection{Classification of interactions and geometrical cross-section}
\label{subsection-basic-quantities}

\begin{figure}
    \parbox{0.59\linewidth}{\centering  \includegraphics*[width=75mm,angle=0]
        {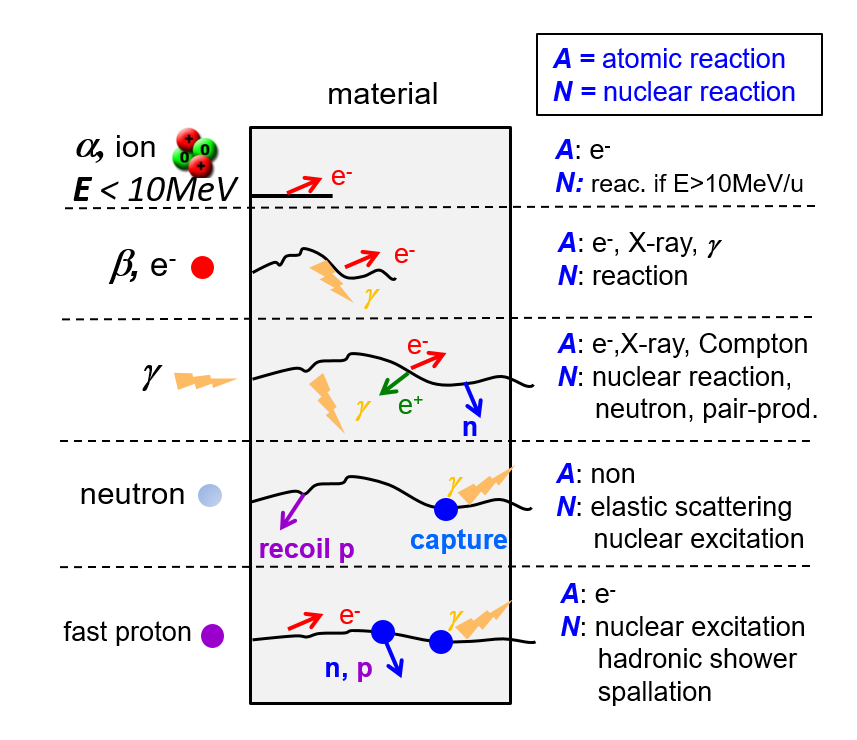} }
    \parbox{0.40\linewidth}{\centering  \includegraphics*[width=55mm,angle=0]
        {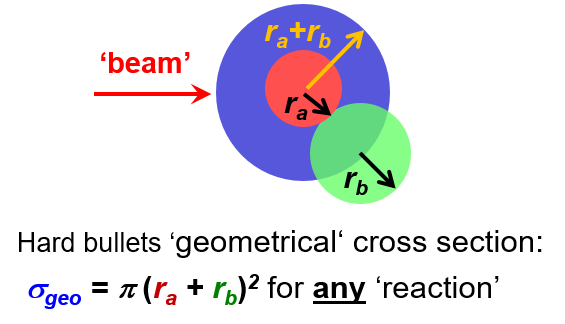} } \vspace*{-3mm}
    \caption{Left: The basic atomic and nuclear interaction of various projectiles with matter. Right: Visualization concerning the geometrical cross-section for the collision of two spherical bullet with radius $r_{a}$ and $r_{b}$, respectively.}
    \label{sigma-geo}
\end{figure}

Particles passing through matter are decelerated by the collision with the target electrons and nuclei. The~interaction can be based on atomic physics processes by Coulomb interaction or nuclear physics processes by strong interaction. Neutrons interact only by strong interaction and have a significantly longer range in matter than protons of the same energy. Electrons and photons interact with target electrons and nuclei only by electro-magnetic interaction. The fundamental interactions are schematically summarized in Fig.~\ref{sigma-geo} (left).

Based on simple geometrical considerations of hard-edge objects as depicted in Fig.~\ref{sigma-geo} (right), one can introduce the geometrical cross-section $\sigma_{geo}$ as a representation of a very probable reaction in case the projectile and target reach the assumed distance.
\begin{itemize}
    \item For atomic processes we consider the Bohr radius of $r_{Bohr}=0.053$ nm as the typical size of an~atom. The cross-section by a point-like projectile is
    \begin{equation}\label{eq-geo-atom}
        \sigma_{geo}^{atom}  = \pi  r_{Bohr}^{2} = 8.8 \cdot 10^{-17} \mbox{~cm}^{2} \simeq  10^{-16} \mbox{~cm}^{2}.
    \end{equation}
    \item For nuclear processes we consider the size to two nuclei with a typical radius of $r_{nucl} = 3$ fm. The~cross-section is for the case of two colliding nuclei
    \begin{equation}\label{eq-geo-nucl}
        \sigma_{geo}^{nucl}  = \pi \left( 2 \cdot r_{nucl} \right) ^{2} = 1.13 \cdot 10^{-24} \mbox{~cm}^{2} \simeq  10^{-24} \mbox{~cm}^{2} \equiv 1  \mbox{~barn}.
    \end{equation}
    The unit barn for the cross-section is introduced. The geometrical cross-section for atomic processes is much larger than for nuclear reactions as $\sigma_{geo}^{atom} \simeq 10^{8} \sigma_{geo}^{nucl}$. Therefore, it is expected that the primary energy loss of charged particle is related to the collision with target electrons; however, the activation of material is determined by nuclear reactions, see Section~\ref{subsection-nuclear-stopping}.
\end{itemize}
Frequently, the mass attenuation or absorption coefficient $\mu$ with the unit [cm$^2$/g] is used. It is related to the cross-section $\sigma$ for the same reaction via Avogadro constant $N_{A}$ and the atomic weight as
\begin{equation}
    \mu =  \frac{N_{A}}{A} \cdot \sigma ~~.
\end{equation}
The mean free path $\lambda_{\mbox{\footnotesize mfp}}$ of a projectile is the average distance between two successive collisions with a~relation to the cross-section by
\begin{equation}\label{maen-free-path}
    \lambda _{\mbox{\footnotesize mfp}} = \frac{1}{n_{t} \sigma} = \frac{M}{\rho N_{A} \sigma},
\end{equation}
where $n_{t}$ is the volume density of scattering centres in the target, $\rho$ the material density, $M$ the mass per mol and $N_{A}$ the Avogadro constant.  For protons as projectile and electrons as target, the energy loss $\frac{\mbox{d}E}{\mbox{d}x}$ per unit of length is a practical measure. The energy loss by multiple scattering (mean free path $\ll$ geometrical size) is given by the integral over the differential cross-section $\frac{\mbox{d}\sigma}{\mbox{d}W}$ as a function of the~energy transfer $W$
\begin{equation}\label{eq-dedx-sigma}
    \frac{\mbox{d}E}{\mbox{d}x}  = n_{t} \int_{W_{min}}^{W_{max}} W ~\frac{\mbox{d}\sigma}{\mbox{d}W}~~\mbox{d}W~~.
\end{equation}
The integration has an upper limit $W_{max}$, e.g. given by kinematic boundaries, see Eq.~(\ref{W-max}) concerning proton-electron scattering. Additionally, there might be a lower integration limit $W_{min}$, e.g. for proton-electron scattering related to the target's atomic bound states or band-gap.

\subsection{Electronic stopping of charged particles}
\label{subsection-electronic-stopping}
For charged projectiles, the energy loss in a material is mainly due to the collision
of the projectile with target electrons, the so-called
electronic stopping power. Due to the different masses of ions and electrons,
the energy transfer to the electron per collision is, typically
below 100 eV, i.e. only a small fraction of the projectile's kinetic energy. The
electronic stopping power $\frac{\mbox{d}E}{\mbox{d}x}$ can be approximated by the~semi-classical Bethe-formula
which is written in its simplest form as \cite{leo-buch,knoll-buch,jackson-buch,ziegler-www,part-data-PhysRevD2018}
\begin{equation}
    -\frac{\mbox{d}E}{\mbox{d}x} = 4 \pi N_{A} r_{e}^{2} m_{e} c^{2} \cdot
    \frac{Z_{t}}{A_{t}} \rho \cdot Z_{p}^{2} \cdot
    \frac{1}{\beta^{2}}
    \left[ \frac{1}{2}
    \ln \frac{2 m_{e} c^{2} \gamma^{2} \beta^{2} W_{max}}{I^{2}} - \beta^{2}
    \right]
    \label{eq-bethe-bloch}
\end{equation}
with the constants: $N_{A}$ the Avogadro number, $m_{e}$ and $r_{e}$ the rest mass
and classical radius of an electron and $c$ the velocity of light. The target
parameters are: $\rho$ density of the target with nuclear mass $A_{t}$ and
nuclear charge $Z_{t}$; the quantity $  n_{e}= \frac{Z_{t}}{A_{t}} \rho $ is
the electron density. $I$ is the mean ionization potential for
removing one electron from the target atoms; a rough approximation for
a target with nuclear charge $Z$ is $I \simeq Z_{t} \cdot 10$ eV, more
precise values are given e.g. in Refs. \cite{ziegler-www,part-data-PhysRevD2018}.
The projectile parameters are: $Z_{p}$ nuclear charge of the ion with velocity
$\beta$ and Lorentz factor $\gamma =(1-\beta^{2})^{-1/2}$. The maximal energy $W_{max}$ transferred from relativistic proton or ion projectile of rest mass $M$ to the target electrons of rest mass $m_{e}$ is given by kinematic considerations for a central collision as \cite{leo-buch,knoll-buch,jackson-buch,ziegler-www,part-data-PhysRevD2018}
\begin{equation}\label{W-max}
    W_{max} =  \frac{2 m_{e} c^{2} \beta^{2} \gamma^{2} }{1+2\gamma m_{e}/M + \left( m_{e}/M\right)^{2}} ~~~~ \mbox{which converges for} M/\gamma \gg m_{e} ~\mbox{to}~~ W_{max}= 2 m_{e} c^{2} \beta^{2} \gamma^{2} ~~.
\end{equation}
The approximation can be interpreted further: For proton or ion impact with non-relativistic energies, i.e. $\gamma \simeq 1$, the velocity of the emitted electron is maximal double of the projectile velocity as expected from classical, mechanical collisions between hard bullets.

The dependence of the Bethe-formula on the square of the target charge $Z_{p}$ is remarkable.  For ionic projectiles, the Bethe-formula is modified since ions travelling through matter are not always bare nuclei, but might have still
some inner electrons and $Z_{p}$ is replaced by an effective charge $Z_{p}^{eff}(\beta)$ in dependence of the velocity $\beta$. One parametrization is $ Z_{p}^{eff}(\beta) \simeq Z_{p} \left[1- \exp\left( -Z_{p}^{-2/3} c\beta /v_{Bohr} \right) \right]$ based on Bohr's atomic model using the Bohr velocity $v_{Bohr}= 0.73\%c$ \cite{ziegler-JAP99}. Better suited parametrizations are included in the semi-empirical codes SRIM
\cite{ziegler-www}, LISE++ \cite{lise++www}, and, as the atomic interaction part of the nuclear physics codes, FLUKA \cite{fluka-www} and GEANT4 \cite{geant4-www}.

\begin{figure}
    \centering \includegraphics*[width=80mm,angle=0]
    {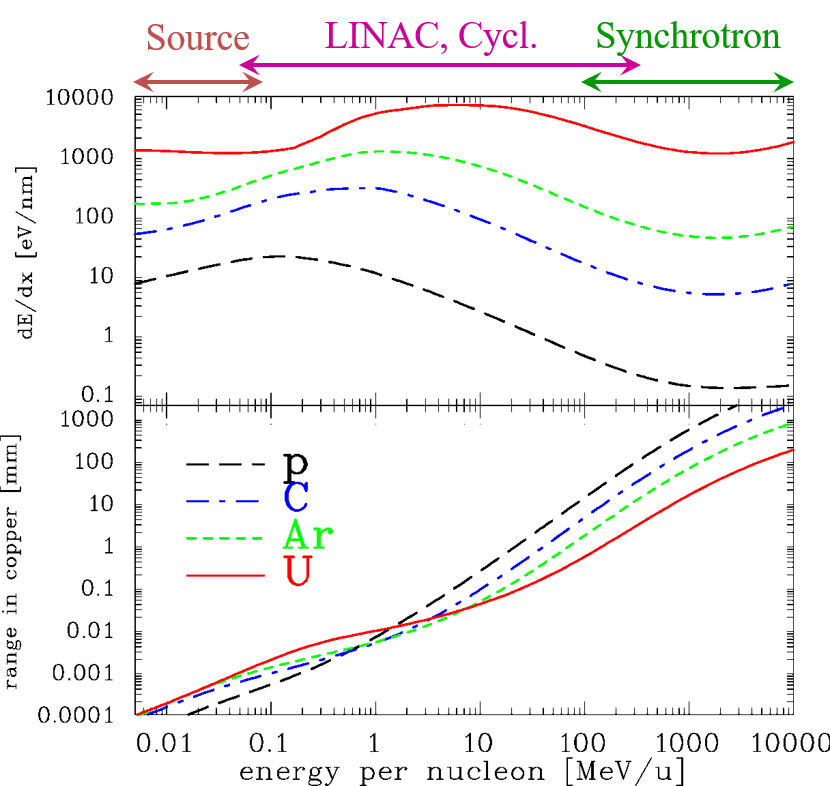} \vspace*{-2mm}
    \caption{The energy loss at the surface
        and the range in copper as a function of the kinetic energy for several
        ions. The energy interval is plotted from 5 keV/u to 10 GeV/u and the range
        from 100 nm to 1 m. The calculation of the~electronic and nuclear stopping
        uses the semi-empirical code SRIM \cite{ziegler-www}. The typical energies of various proton and ion accelerator types are indicated.}
    \label{copper_range}
\end{figure}

The result of such semi-empirical computation for the energy loss by SRIM is depicted in
Fig.~\ref{copper_range} for different
ions into copper as a representative for a medium-heavy metal target. The energy loss is maximal for ions with kinetic energy
around 100 keV/u to 7 MeV/u (corresponding to velocities $\beta \sim 1.5$~\%
to $ 12$ \%) depending on the ion species. These are typical energies of a
proton or ion LINAC. For relativistic
energies above 1 GeV/u, the energy loss is nearly constant; these are typical
energies of particles extracted from a synchrotron.
Note that the Bethe-formula is valid for any charged projectile and depends on the projectile velocity $\beta$.

The range $R$ of the projectile of kinetic energy $E_{kin}$ is calculated numerically via
\begin{equation}
    R =\int_{0}^{E_{kin}} \left( \frac{\mbox{d}E}{\mbox{d}x} \right)^{-1} \mbox{d}E
    \label{eq-range-calc}
\end{equation}
and has an approximately scaling for ions
above $\simeq 10 $ MeV/u \cite{leo-buch}
\begin{equation}
    R \propto E_{kin}^{1.75} ~.
    \label{eq-range-bethe-bloch}
\end{equation}
The results are shown in Fig.~\ref{copper_range}. For kinetic energies below some 10 MeV/u for protons or ions are shorter than about 1 mm, and a beam particle is stopped in the vacuum tube. But even for higher energies to about several 100 MeV/u most beam particles are stopped in the vacuum chamber as it collides in most cases under a grazing incidence. As discussed in Section \ref{subsection-nuclear-stopping}, nuclear reactions might be possible leading to isotopes with lifetimes above several minutes and, consequently, to an activation.

Due to the interaction's statistical nature, fluctuations in the energy transfer occur, the resulting variation of the projectile’s trajectories is called straggling. Related to the significant mass difference between projectile and target electrons, the longitudinal and transversal straggling is low compared to the range; analytic approximations are discussed in Refs. \cite{leo-buch,knoll-buch,jackson-buch, ziegler-www,part-data-PhysRevD2018}. Therefore, the projectile path is almost straight with a defined range; the actual properties are calculated by Monte-Carlo methods.

For energies below  $E_{kin} < 10$ keV/u, e.g. at the end of a projectile trajectory, the probability of Coulomb collisions between the projectile and target nucleus dominates the stopping power \cite{leo-buch,ziegler-www}; this is called nuclear stopping (note that the Coulomb force mediates and not the strong interaction). The energy transfer in this single collision within a crystal material can cause a knock-out of the target nucleus from its lattice position resulting in a displacement. This process leads to radiation damage of materials as discussed e.g. in Refs. \cite{Mokhov-CAS2016,Kiselev-CAS2013,bertarelli-CAS2016}.

\subsection{Effects related to material heating}
\label{subsection-material-heating}

\begin{figure}
    \parbox{0.45\linewidth}{\centering  \includegraphics*[width=68mm,angle=0]
        {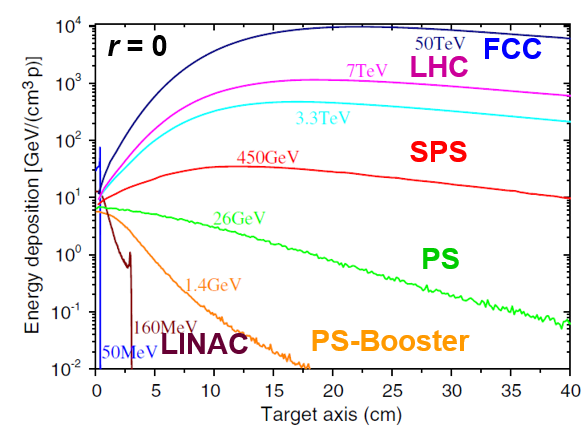} }
    \parbox{0.54\linewidth}{\centering  \hspace*{-5mm} \includegraphics*[width=88mm,angle=0]
        {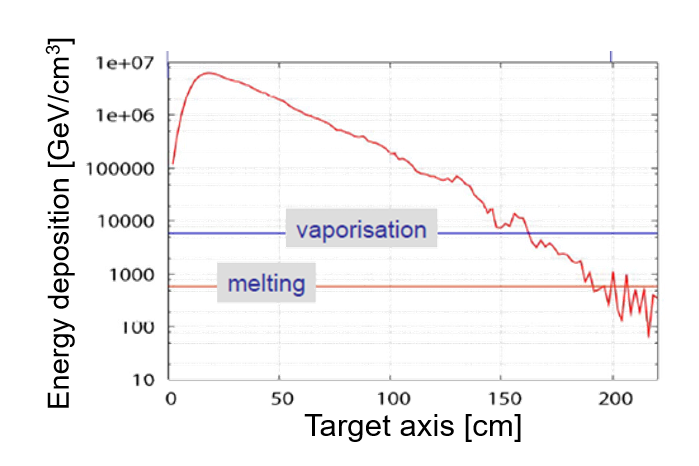} } \vspace*{-3mm}
    \caption{Left: The energy deposition per incident proton on the central axis $r=0$ of a various beam energies and an~rms beam size is $\sigma = 0.2$ mm, the abbreviations refer to the CERN accelerators; reproduced from Ref. \cite{nie-prab2017}.  Right: The~specific energy deposition along the beam axis estimated for a filling of LHC with 2808 bunches each of  $1.1\cdot 10^{11}$ protons at 7 TeV and rms beam size of $\sigma = 0.2$ mm; the total beam energy is 350 MJ, the thresholds for melting and vaporization are indicated; reproduced from Refs. \cite{Schmidt-CAS2014, tahir-jap2005}.}
    \label{energy-deposition-cenrtal-cern}
\end{figure}

Related to the electronic stopping of charged projectiles, as given by Bethe-formula of Eq.~(\ref{eq-bethe-bloch}), electrons are knocked out and penetrate through the material. Within a short range below 100 $\mu$m, they distribute kinetic energy to other electrons and excite phonons; the latter is directly related to the material temperature. As a result, the local temperature close to the projectile track is increased. Examples for such a~calculation are depicted in Fig.~\ref{energy-deposition-cenrtal-cern} for a proton beam ranging from 50 MeV (the energy of LINAC2 at CERN) and 50 TeV (the energy of protons in the proposed collider FCC) impacted in solid copper \cite{nie-prab2017}. The energy is locally deposed close to the proton track and shown along the beam path centre. For the~low energy case, a distinct maximum is reached at a depth corresponding to the $ \simeq 4$ mm range of the~projectiles, see Fig.~\ref{copper_range}. For higher proton energies, the maximum is reached inside the copper block with distances of several cm, leading to heating of the material for the inner part. To calculate the temperature increase $\Delta T$ the volumetric energy loss per projectile $\frac{\mbox{d}E}{\mbox{d}V} $ has to be multiplied with the number of beam particles $N$, divided by the density $\rho$, and the potentially temperature-dependent heat capacitance $c(T)$ as
\begin{equation}\label{eq-temperure-increase}
    \Delta T = N \cdot \frac{\mbox{d}E}{\mbox{d}V}  \cdot \frac{1}{\rho \cdot c(T)} ~~.
\end{equation}
Figure \ref{energy-deposition-cenrtal-cern} depicts such a simulation of the immediate impact of a complete filling of LHC at top energy and the nominal beam size of $\sigma_{\text{LHC}} = 0.2$ mm. This acts as a worst-case scenario where the entire beam from LHC is lost and impacts a beam dump without further defocusing. In this case, the beam dump would melt and even evaporating would take place inside the copper absorber, leading to a pressure increase, stress, swelling, and irreversible material deformation leading to holes in the dump in extreme cases. For the depicted example, the beam impact duration is related to one revolution period in LHC of 86 $\mu$s. During this short period, the heat dissipation related to the heat conductivity can be neglected. The physical basis and measurements are discussed in Refs. \cite{Mokhov-CAS2016,bertarelli-CAS2016,nie-prab2017, tahir-jap2005,kain-pac2005,nie-prab2019}.

\begin{figure}
    \centering \includegraphics*[width=160mm,angle=0]
    {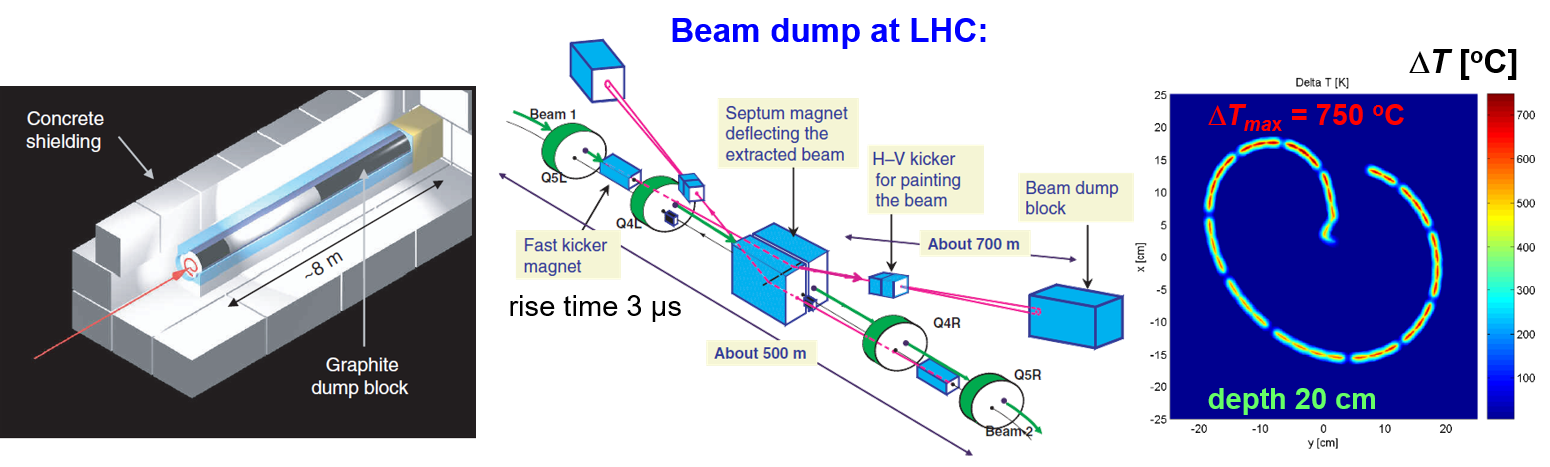}
    \caption{Left: The geometrical layout of the LHC beam dump made of 8 m long graphite block of $\oslash 0.7$ m diameter. Middle:  Schematic drawing of the kicker magnet inside LHC and the horizontal and vertical kickers in the transfer line to the dump. Right: The calculation of the temperature increase for the beam distribution on a $50 \times 50$ cm$^{2}$ area and a depth of 20 cm; all figures reproduced from Ref. \cite{Schmidt-NJP2006}.}
    \label{Beam-dumpLHC-all}
\end{figure}

As an example, we briefly discuss the LHC beam dump layout as depicted in Fig.~\ref{Beam-dumpLHC-all}. The beam dump itself is made of graphite with a length of 7 m to stop the 8 TeV protons. The low density of graphite enables at relatively low stopping power due to the scaling $\frac{\mbox{d}E}{\mbox{d}x} \propto \rho$ and high sublimation temperature for the graphite. The surrounding concrete is required to absorb the charged particles for nuclear reactions and to moderate the neutrons. The beam can be kicked out from the LHC by magnetic kickers with a~rise time of about 3 $\mu$s. The 700 m long transfer line contains dipole magnets that change their kick angle during the one-turn extraction time of 86 $\mu$s to paint the beam on the beam dump. Moreover, the~beam size increases significantly to $\sigma_{dump} \simeq  2~\mbox{mm}\simeq  10 \cdot \sigma_{\text{LHC}}$ mm. The resulting beam distribution is depicted in Fig.~\ref{Beam-dumpLHC-all} (right). It shows a calculation of the temperature increase within the graphite block for a depth of 20 cm close to the maximal energy deposition. For nominal beam parameters, the maximum temperature increase is limited to $\Delta T_{max} \simeq 750~^{o}$C and therefore within the safety margin for regular operation.

For a beam dump downstream of a LINAC or cyclotron, the exposure time can be significantly longer than for the discussed LHC case of 86 $\mu$s. For such an operational mode, dissipation by heat conduction are of relevance leading to a partial differential equation for the temperature change in the~form
\begin{equation}
    \frac{\partial T (\vec{x},t)}{\partial t} = \frac{\lambda}{\rho c(T)}
    \mbox{div~} \mbox{grad~} T(\vec{x},t) + \frac{N}{\rho c(T)} \cdot \frac{\mbox{d}E(\vec{x},t)}{\mbox{d}V}
\end{equation}
for the temperature $T$ as a function of time $t$ at each position $\vec{x}$ inside the material, $\lambda$ is the heat conductivity, $\rho$ the density and $c(T)$ the temperature dependent heat capacitance. The source term $ \left( N\frac{\mbox{d}E(\vec{x},t)}{\mbox{d}V} \right)$ is the volumetric power deposition by the beam of $N$ particles. The dynamical material response including heat- and pressure-related effects is described in Ref. \cite{bertarelli-CAS2016}.

Besides the thermal properties of the entire beam dumped to a target, the thermal effect related to partial beam losses is even more relevant. The beam loss is caused by a device malfunction, improper operation settings, or beam halo development. The beam particles might impinge on the vacuum chamber under a grazing angle, leading to a penetration length significantly longer than the vacuum chamber thickness. For sufficient high beam energies, the beam particles are transmitted and stopped in external devices like magnets. In particular, for superconducting magnet coils, the related energy deposition can heat the coils above the critical temperature and consequently causes a quench, i.e. the loss of superconductivity. The energy loss and related heating might lead to material damage, which is calculated in the same manner as above. However, the heat capacitance of cold material decreases significantly with temperature \cite{duthil-cas2013}: For insulators, the scaling is $c_{\text{phonon}} (T) \propto T^{3}$ as the heat is distributed on lattice vibrations (phonons). For normal-conducting metal, the thermal properties of electrons contribute additionally with a scaling of $c_{\text{metal}}(T) \propto \alpha T + \beta T^{3}$. For superconductors an exponential scaling applies $c_{\text{sc}}(T) \propto T_{C} \cdot e^{(-\gamma T_{c}/T)}$ with $T_{C}$ is the critical temperature and $\alpha, \beta$ and $\gamma$ are material constants. For all materials, the heat capacitance is lower than at room temperature, which increases the hazard and potential accelerator downtime, e.g. required for the magnet recovery after a quench; the physics of quenches and mitigations are discussed in Refs. \cite{many-cas2013,mess-cas88} and tests are described in Ref. \cite{auch-prab2015}. The same risk applies to superconducting rf-cavities and insertion device at light sources. The maximal allowed particle impact and related temperature increase is calculated numerically and used as thresholds for a~machine protection system, see Section~\ref{chapter-machine-protection}.

\subsection{Interaction of electrons}
\label{subsection-electron-stopping}

\begin{figure}
    \parbox{0.59\linewidth}{\centering  \includegraphics*[width=80mm,angle=0]
        {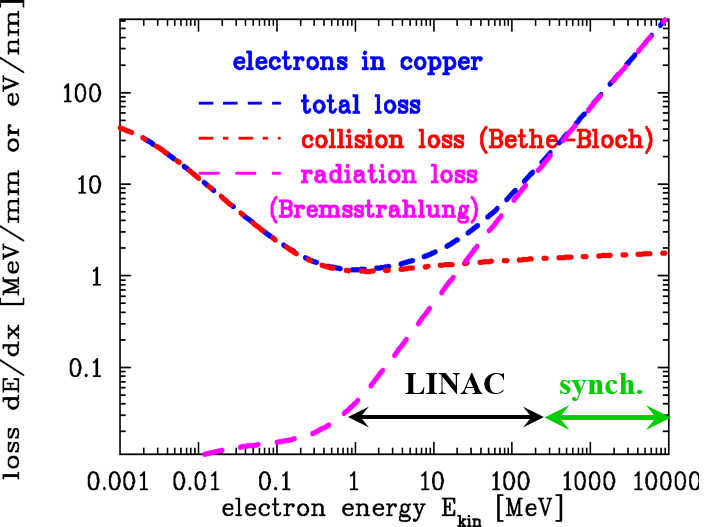} }
    \parbox{0.40\linewidth}{\centering
        \includegraphics*[width=40mm,angle=0]{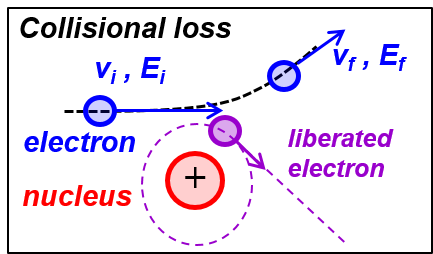} \\ \vspace*{5mm}
        \includegraphics*[width=40mm,angle=0]{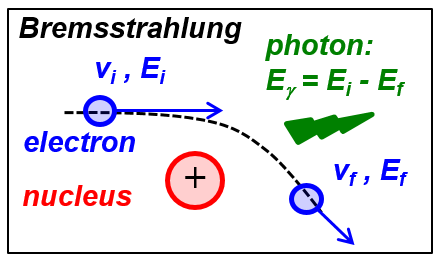} }
    \caption{Left: Energy loss of electrons in copper showing the two regimes for collisional loss and Bremsstrahlung, data calculated by \cite{berg_NIST}. Right: Schematics for collisional loss and Bremsstrahlung.}
    \label{electron-energyloss}
\end{figure}

The stopping of electrons in matter differs from protons and ions as the sum of the two relevant processes of collisional loss $\mbox{d}E/\mbox{d}x|_{col}$ and radiation loss $\mbox{d}E/\mbox{d}x|_{rad}$ are added, see Fig.~\ref{electron-energyloss}. A modified Bethe-formula describes the collisional loss for electrons due to electronic stopping $\mbox{d}E/\mbox{d}x|_{col}$ \cite{leo-buch, part-data-PhysRevD2018}. This regime dominates for energies below 1 MeV as corresponding to a velocity $\beta = v/c= 94$ \% and Lorentz-factor $\gamma =2.96$. Even small electron accelerators reach energies significantly above 1 MeV.  The trajectories of primary electrons in the target are more curved than for ions due to the potential high energy- and momentum transfer in a single collision; therefore, electrons have much larger lateral and longitudinal straggling than protons.

For energies above a few 10 MeV the radiation loss by Bremsstrahlung dominates, i.e. the emission of photons by acceleration or deceleration
of an electron in the vicinity of the target nucleus, see e.g. Refs. \cite{Lechner-CAS2017,leo-buch,part-data-PhysRevD2018}. The energy loss can be approximated for an electron with rest mass $m_{0}$ and kinetic energy $E_{kin} \gg m_{0}c^{2}$ in a target of number density $n_{t}$ and charge $Z_{t}$ by Refs. \cite{leo-buch,knoll-buch}
\begin{equation}\label{eq-bremsstrahlung}
    \frac{\mbox{d}E}{\mbox{d}x} |_{rad} \simeq \frac{e^{4}  E_{kin} n_{t} Z_{t} \left( Z_{t} +1 \right)}{137m_{0}^{2}c^{4}} \left(4 \ln \frac{2E_{kin}}{m_{0}c^{2}} -\frac{4}{3}  \right) ~~.
\end{equation}
This radiation loss scales roughly linear to the electron energy and quadratic to the target charge $Z_t$. The strong dependence on the target charge is expected as the electron's deceleration causes the photon emission. The generated Bremsstrahlung photons at typical accelerators have energies much above 1~MeV. In the vicinity of a nucleus, these photons can create electron-positron pair, which themself can have significant kinetic energy; see Section~\ref{subsection-photon-stopping}. These second-generation electrons or positrons can emit again Bremsstrahlung photons if their energy is sufficient; this leads to an electron-magnetic shower with multiple emission of $\gamma$-rays initialized by a single, high energetic electron. Moreover, nuclear processes are possible: Bremsstrahlung photons can excite the nuclei in material to a collective mode, the so-called giant resonances. The decay often proceeds via a neutron emission; this is discussed in Section \ref{subsection-photon-stopping}. Due to this process, neutrons are produced at electron accelerators, and a relative thick concrete shielding of the beam-line is required. In addition, these neutrons can initialize a nuclear reaction leading to activation of components at electron accelerators.

\subsection{Interaction of photons}
\label{subsection-photon-stopping}
At electron accelerators, Bremsstrahlung photons are produced due to the electrons' deceleration in matter, see Section \ref{subsection-electron-stopping}. Related to the electronic stopping of protons and electrons, the target atom is excited, or electrons are liberated from the atom leading to fluorescence photons up to the hard X-ray region of several 100 keV. Moreover, at proton and ion accelerators, $\gamma$-rays up to typically 10 MeV are emitted as a nuclear reaction product by proton or ion impact, see Section \ref{subsection-nuclear-stopping}. The absorption of photons is one design consideration of the concrete shield around an accelerator and related to the Machine Protection System, see Section~\ref{subsection-MPS-hardware}.

\begin{figure}
    \parbox{0.99\linewidth}{\centering  \includegraphics*[width=155mm,angle=0]
        {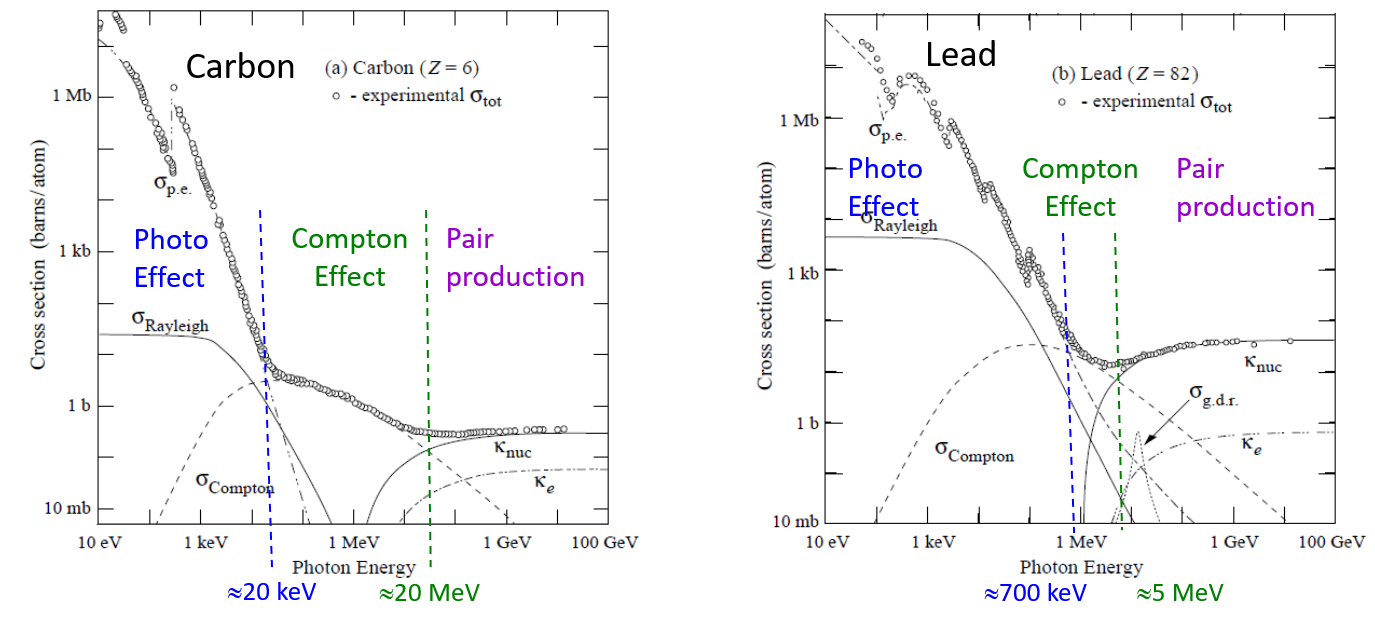} }
    \caption{Total cross-section for photon scattering in carbon (left) and lead (right). The contributions of the individual processes are indicated with the following abbreviations: $\sigma_{\text{p,e}}$ photon electron liberation, $\kappa_{\text{nuc}}$ pair production in the~field of a nucleus, $\kappa_{\text{e}}$ pair production in the~field of a nucleus,  $\sigma_{\text{g.d.r.}}$ giant dipole resonance leading to the~emission of a neutron. Note the large logarithm scaling; figure reproduced from Ref. \cite{part-data-PhysRevD2018}, data also available at Refs. \cite{berg_photon_NIST,ENDF-wwww}.}
    \label{photon-crosssection}
\end{figure}

\begin{figure}
    \parbox{0.49\linewidth}{\centering  \includegraphics*[width=50mm,angle=0]
        {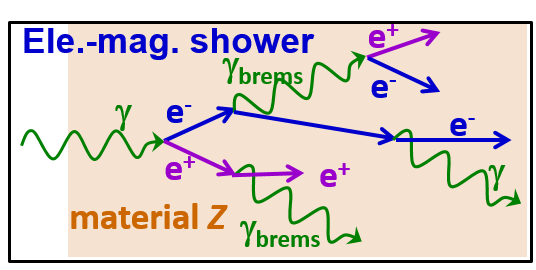} }
    \parbox{0.49\linewidth}{\centering
        \includegraphics*[width=40mm,angle=0]{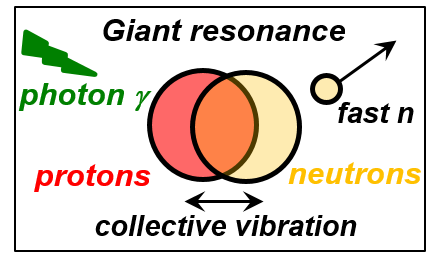}}
    \caption{Left: Scheme of an electro-magnetic shower. Right: Scheme of a giant dipole resonance of type ($\gamma$,n).}
    \label{scheme-giant-resonance}
\end{figure}

Figure~\ref{photon-crosssection} depicts the cross-section for important photo absorption processes for a light and heavy target. The various photon interactions have characteristic energy dependence and scaling with the target composition \cite{Lechner-CAS2017,part-data-PhysRevD2018,leo-buch}. The total cross-section is the sum of the following processes:
\begin{itemize}
    \item \textit{Rayleigh scattering, reaction} {$ \gamma + \mbox{atom} \rightarrow \gamma + \mbox{atom}$}:\\
    The elastic scattering of photon by the target atom is described by Rayleigh Scattering. As this process does not change the photon energy, it does not contribute to the absorption. The cross-section scales approximately as $\sigma_{\text{Ray}} \propto Z_{t}^{2}$ with the target nuclear charge $Z_{t}$.
    \item \textit{Photo effect, reaction} {$ \gamma +\mbox{atom} \rightarrow \mbox{atom}^+ + e^{-}$}:\\
    The photon is absorbed and an electron is emitted. The cross-section shows some steps whenever the energy is sufficient to emit an inner shell electron. The overall cross-section has a strong dependence on the nuclear charge $Z_{t}$ as it scales approximately $\sigma_{\text{p,e}} \propto Z_{t}^{4}$. This process dominates the low energy photon absorption.
    \item  \textit{Compton effect, reaction} {$ \gamma +\mbox{atom} \rightarrow \gamma' + \mbox{atom}^+ + e^{-}$}:\\
    The incoming photon is inelastically scattered at an atom, realizes some of the energy to liberate an~electron and a photon $\gamma '$ is emitted. The cross-section scales approximately as $\sigma_{\text{Com}}  \propto Z_{t}$.  Within an energy range of typically 10 keV to 10 MeV, the cross-section for Compton effects dominates.
    \item \textit{Pair production, reaction} {$ \gamma +\mbox{nucleus} \rightarrow e^{-} + e^{+} + \mbox{nucleus}$}:\\
    A photon of energy above 1022 keV, twice the rest mass on an electron, can create an electron-positron pair. To conserve energy and momentum, pair production is only possible in the vicinity of a nucleus with an approximated scaling $\sigma_{\text{pair}}  \propto Z_{t}^{2}$. The cross-section dominates for energies above 10 MeV. If the electrons and positrons' kinetic energy is sufficiently high, Bremsstrahlung photons will be emitted, leading to a second generation electron-positron pair. In case of further multiplication, the process is called electro-magnetic shower \cite{Lechner-CAS2017,part-data-PhysRevD2018,leo-buch}, see Fig.~\ref{scheme-giant-resonance}.
    \item \textit{Photon absorption via nuclear giant resonance, reaction} {$ \gamma +^{\mbox{A}}\mbox{nucleus} \rightarrow n + ^{\mbox{A-1}}\mbox{nucleus}$}: \hfill \\
    The $\gamma$ can be absorbed by a so-called nuclear giant resonance due to its large cross-section for heavy targets up to almost the nuclear geometrical cross-section of 1 barn defined by Eq.~(\ref{eq-geo-nucl}). As depicted in Fig.~\ref{scheme-giant-resonance}, the photon excites a collective motion of protons versus neutrons, resulting in an oscillating electrical dipole moment \cite{povh-book}. The nucleus de-excites in many cases via neutron emission, i.e. reaction $^{A}X(\gamma,n)^{A-1}X$. This nuclear reaction is the source of neutrons at electron accelerators excited as the second step by Bremstrahlung photons. It leads to the shielding requirement at electron accelerators and activation by the neutrons' nuclear reaction.
\end{itemize}

\subsection{Nuclear physics processes for proton and ion impact}
\label{subsection-nuclear-stopping}

\begin{figure}
    \centering  \includegraphics*[width=120mm,angle=0]
    {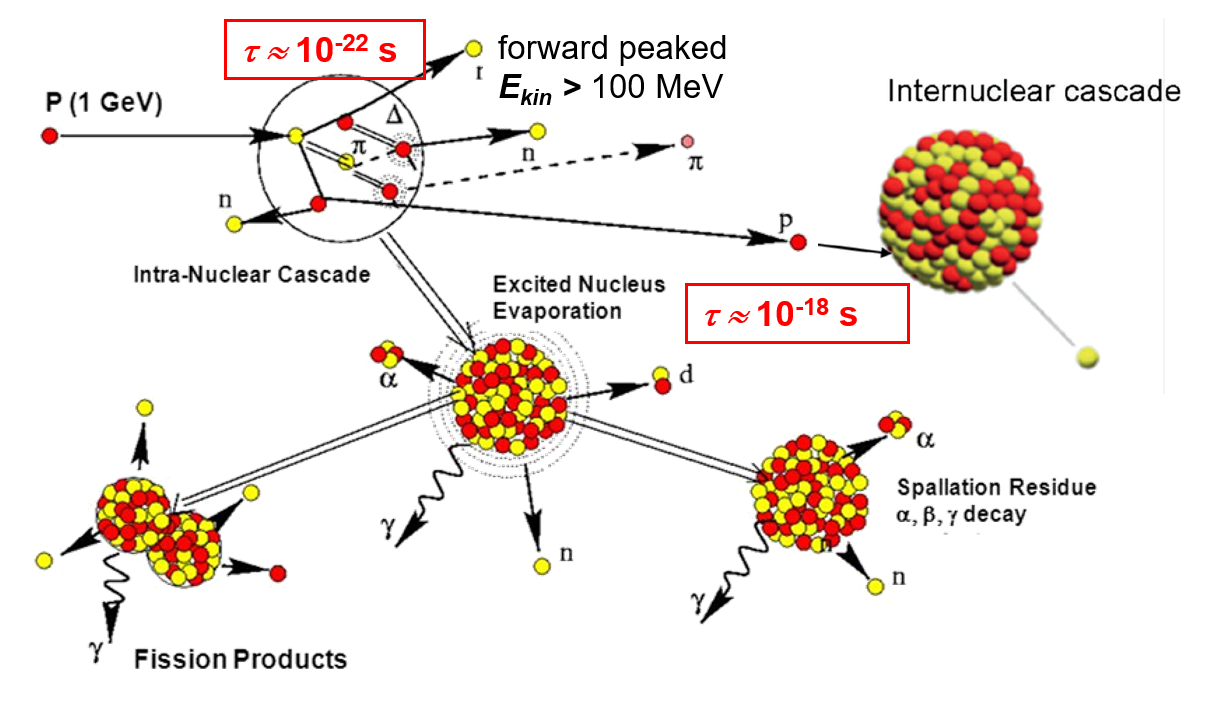} \vspace*{-3mm}
    \caption{The reaction product by the spallation of protons of 1 GeV energy on a heavy target, reproduced from \cite{Kiselev-CAS2013}.}
    \label{spallation-all}
\end{figure}

The strong interaction between the projectile proton or ion and the target nuclei can lead to a wide variety of emitted particles and radioactive nuclei, resulting in long-term activation of the accelerator components. Exemplarily, we consider the impact of a relativistic proton with energy of 1 GeV on a~medium-heavy (e.g. iron or copper) or heavy target (e.g. tungsten or lead). The relevant processes for this so-called spallation reaction are depicted in Fig.~\ref{spallation-all} for a central collision. Using some simplifications, several steps can be distinguished:
\begin{itemize}
    \item \textit{Intranuclear cascade:} The projectile interacts with the individual nucleons, excites quark-state based resonances, like $\Delta$-resonances within the proton, or create mesons, most probable pions. Within $\tau \simeq 10^{-22}$~s it leads to the emission of fast pions, neutrons or protons with kinetic energies above $E_{kin} > 100$~MeV.
    \item \textit{Evaporation process:} The excitation energy is dissipated between the nucleons. The process leads to an excitation of nuclear levels followed by the emission of light nuclei, like $\alpha$-nuclei, deuterons or the energy is released by neutron or $\gamma$ emission. The kinetic energy is several MeV, i.e. lower than during the preceding intranuclear cascade. The duration for the energy dissipation is about  $\tau \simeq 10^{-18}$ s.
    \item  \textit{Decay of unstable isotopes:} In many cases, the first two steps of the spallation lead to unstable isotopes, but those can have long lifetimes (in the sense of nuclear physics, e.g. 1 ms is regarded as a long duration). The energy is dissipated between all nucleons, and the nucleus decays via  $\alpha$, $\beta$ decay or the emission of $\gamma$-rays or neutrons. For a target of heavy elements, fission might also be a prominent decay channel. The typical energy of the emitted particles are in the order of MeV; the duration can be up to  $\tau \simeq 1$ ms.
    \item \textit{Internuclear cascade:} An alternative channel to the evaporation process is the dissipation of the~collisional energy within the nucleus, which might result in the emission of a proton or neutron.
    \item \textit{Activation:} after the above steps, the final nuclei might be unstable with long lifetimes much above $\tau \gg 1$ s. It leads to the activation of accelerator components after the beam shut-off. The related activation might hinder persons' access for a duration from hours to several months, depending on the cocktail of produced isotopes.
\end{itemize}
The cross-section of the above-described processes depends significantly on the projectile and target isotopes, and no scaling laws are given here; remarkably, those interactions can emit several neutrons. A~more detailed discussion can be found e.g. in Refs. \cite{Mokhov-CAS2016,Kiselev-CAS2013,protection-part-data-PhysRevD2018,tho-handbook,sull-buch, shultis-buch,ncrp-report} and in textbooks on nuclear physics \cite{povh-book}.

\begin{figure}
    \centering  \includegraphics*[width=120mm,angle=0]
    {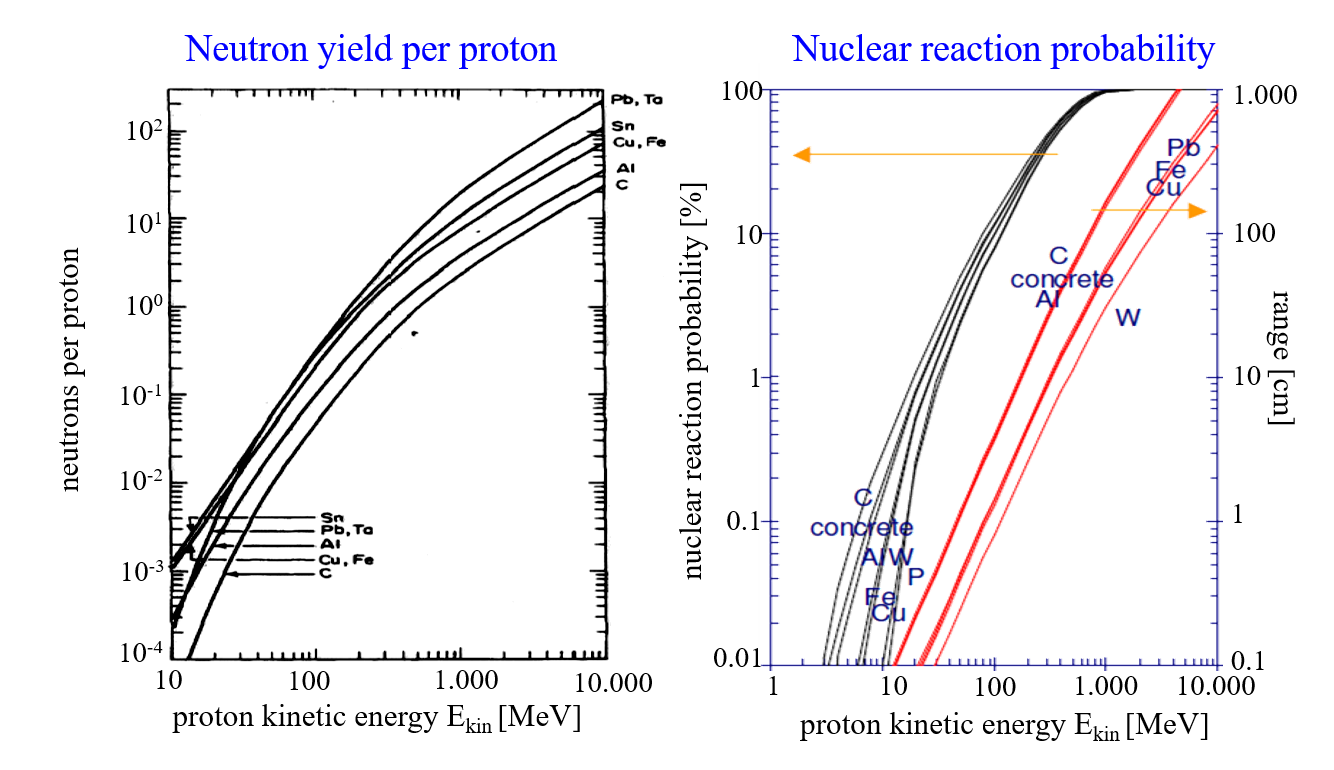} \vspace*{-4mm}
    \caption{Left: The amount of neutrons per proton impact as a function of proton energy. Right: The probability of any nuclear reaction by protons as a function of energy. Both plots concerns the impact on a thick target, reproduced from Ref. \cite{tho-handbook}.}
    \label{reaction probability-proton-energy}
\end{figure}

For the design of the accelerator's shielding, the neutron yield is of particular interest. Figure~\ref{reaction probability-proton-energy} shows the neutron yield per incident proton.  Protons with energies above 1 GeV produce more than one neutron per incident particle in all materials. The number of neutrons scales approximately as $N_{n} \propto E_{kin}^{2}$ from 10 MeV to 1 GeV. Moreover, the figure shows the calculated number of any nuclear reaction for proton impact. Several nuclear reactions are expected for energies above 1 GeV; most of these reactions lead to a long lifetime isotope and a related activation. The figure supports the statement that nuclear reactions are very probable for proton impact above 100 MeV and show that activation and the required shielding are of great relevance for proton and ion accelerators. The depicted reaction probability concerns the collisions inside a thick target. The phrase 'thick target' is defined for a proton penetration path length comparable to the range given by the electronic stopping power of Eq.~(\ref{eq-range-bethe-bloch}) and Fig.~\ref{copper_range}. Additionally, the ranges of various materials are depicted for comparison in Fig.~\ref{reaction probability-proton-energy}. For ion impact, further reactions are possible, e.g. the projectile's nuclear excitation followed by an in-fight emission of particles, leading, e.g. to relativistic neutrons.

As several reaction channels are possible, Monte-Carlo methods are used for the calculation; frequently used codes are FLUKA, MARS, MCNPX and PHITS \cite{fluka-www,mars15-www,mcnpx-www,phits-www} which contains all relevant nuclear excitations and reactions as well as the electronic stopping of charged particle and the attenuation of neutrons; a comparison of various codes and the related physics can be found, e.g. in Refs. \cite{Mokhov-CAS2016,Kiselev-CAS2013,protection-part-data-PhysRevD2018,tho-handbook}.

\begin{figure}
    \parbox{0.49\linewidth}{
        \centering  \includegraphics*[width=85mm,angle=0]
        {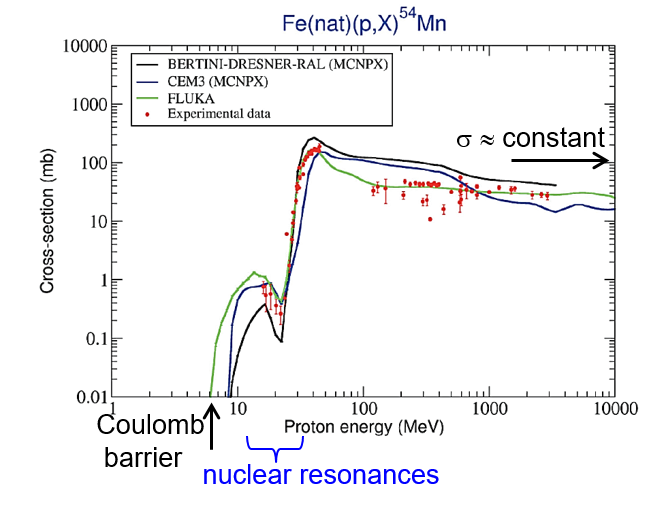}
    }
    \parbox{0.49\linewidth}{\centering  \includegraphics*[width=70mm,angle=0]
        {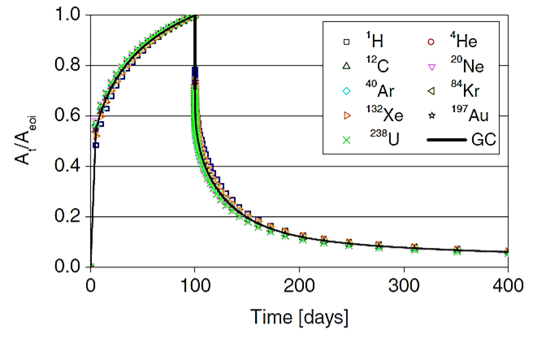} } \vspace*{-3mm}
    \caption{Left: Energy dependent cross-section for the proton induced nuclear reaction of natural iron into $^{54}$Mn. Several models are compared to experimental data, reproduced from \cite{Kiselev-CAS2013}. Right: Time evolution of the normalized activation induced by beam impact of various ions at 500 MeV/u on stainless steel during 100 days and the related decay of the activation, reproduced from Ref. \cite{strasik-prab2010}.}
    \label{iron-nuclear-reactions}
\end{figure}

In most cases, a beam particle collides with the vacuum chamber under grazing incidence, and, if transmitted, it might be stopped in the yoke of a magnet. As a characteristic example, the cross-section of proton impact on natural iron (6\% $^{54}$Fe, 92\% $^{56}$Fe, 2\% $^{57}$Fe) is depicted Fig.\ref{iron-nuclear-reactions} (left) related to the production of a long lifetime isotope and related to activation \cite{Kiselev-CAS2013}. Minimal energy is needed for the reaction as the proton has to overcome the Coulomb barrier of the iron nucleus between 3 and 10 MeV for most projectile-target combinations. The following energy interval up to about 50 MeV is determined by the nuclear excitation where the proton is captured to excited levels. The cross-section comprises resonances; their position and shape depend strongly on the target isotope. Above 100 MeV, the cross-section is almost constant. The production cross-section for the dedicated isotope $^{54}$Mn is a~few percentages of the geometrical cross-section defined by  Eq.~(\ref{eq-geo-nucl}), and this isotope is produced with high probability and lead to activation. The lifetime of $^{54}$Mn  is $\tau = 216$ days and it decay via two $\gamma$ emission of $\simeq$ MeV energy.

The example for proton-induced isotope production shows the relevance for nuclear reactions to the activation of accelerator components and the related delayed access of persons to the area, e.g. in case of maintenance work during the accelerator shut-down. One goal for a machine protection system is to warn the accelerator operators or prevent beam delivery if an extended loss is detected. Figure~\ref{iron-nuclear-reactions} (right) shows an example of the related simulations of the increase of activation during 100 days of permanent beam loss at a given location and the 'cool down' of the activation after beam production~\cite{strasik-prab2010}. Due to the~direct proton-induced reactions or due to daughter nuclei, isotopes with a lifetime between days and years are produced. The cocktail of isotopes is almost independent of the beam ion. The~resulting activation leads to a 'cool down' duration in the order of several 10 days; the composition of the isotope cocktail might change during this time related to the different lifetimes of the radioactive isotopes resulting in a non-exponential decay as, e.g. discussed in Refs. \cite{Kiselev-CAS2013,strasik-prab2010}. Depending on the beam loss and actual activation, the person access is regulated, and maintenance work might be delayed correspondingly.

\subsection{Interaction of neutrons}
\label{subsection-neutron-stopping}
\begin{figure}
    \parbox{0.82\linewidth}{\centering  \includegraphics*[width=132mm,angle=0]
        {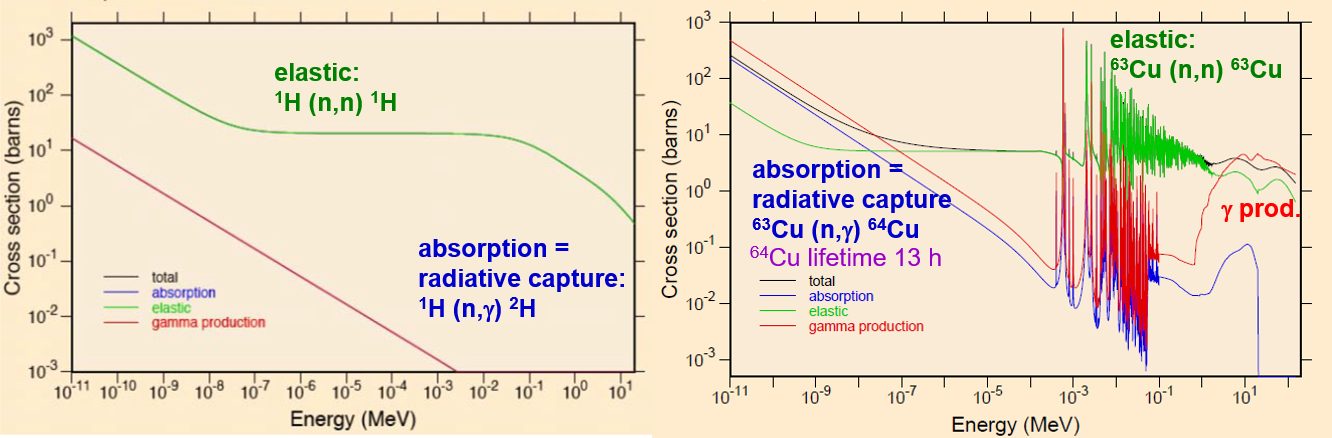} }
    \parbox{0.17\linewidth}{\centering
        \includegraphics*[width=25mm,angle=0]{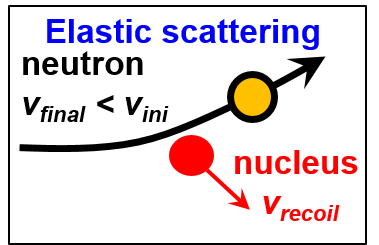} \\ \vspace*{2mm} \hspace*{0.5mm}
        \includegraphics*[width=27mm,angle=0]{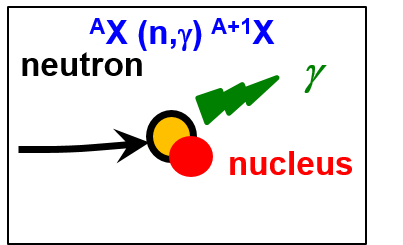} }
    \caption{Cross-section of neutron interactions in hydrogen (left) and copper (centre) calculated by the code ENDF and reproduced from \cite{ENDF-wwww}, schematics for elastic scattering and neutron capture (right).}
    \label{neutron-sigma-hydrogen-copper}
\end{figure}

Neutrons are subject to strong interaction only, but contrary to protons, they do not have to overcome the~Coulomb barrier (about 3 to 10 MeV); therefore, neutrons at thermal velocities interact with the~target nucleus. Two examples for the energy-dependent cross-section of neutrons are shown in Fig.~\ref{neutron-sigma-hydrogen-copper} for an extensive energy range from sub-thermal energies up to several 100 MeV calculated by the code ENDF/B \cite{ENDF-wwww}. As a representative for accelerator components like vacuum pipes, magnets and rf cavities, we first discuss the properties for scattering at a target of a medium-heavy element:
\begin{itemize}
    \item \textit{Elastic scattering:}
    The elastic cross-section describes the neutron scattering at a target nucleus $X$, i.e. the reaction $X(n,n')X$. For sub-thermal energies, the cross-section is even above the~nuclear geometrical value (see Eq.~(\ref{eq-geo-nucl})) as related to the large de Broglie wavelength of the neutron. The~neutron loses some of its kinetic energy as the nucleus overtakes the recoil momentum. In the~energy range from about 10 keV to 10 MeV, the neutron is captured to nuclear single-particle levels leading to a resonance structure for the cross-section with narrow lines; at higher collision energies.
    \item \textit{Absorption:}
    A second branch is related to the absorption of a neutron by the target nucleus $X$, typically as $^{A}X(n,\gamma)^{A+1}Y$. It is a resonance process related to the nuclear levels of the daughter nucleus $Y$, and in many cases, a $\gamma$ is emitted to reach the ground state of the daughter nucleus by electro-magnetic relaxation. At energies above several 100 MeV, the neutron cross-section is comparable to the proton cross-section as both nucleons trigger similar reactions.  By this neutron absorption, radioactive isotopes can be produced, which are subject to $\beta$-decay. This is one source of activation at electron accelerators. At proton accelerators, this is an additional contribution to the activation following the spallation reaction.
\end{itemize}

The second example in Fig.~\ref{neutron-sigma-hydrogen-copper} is the cross-section for scattering on hydrogen: Here, the absorption cross-section is negligible as the neutron binding energy of a deuteron is relatively low (about 2.2 MeV for deuteron, compared to the binding energy per nucleon for medium-heavy elements of about 8 MeV). However, the elastic process is very probable. Due to the almost equal mass of neutron and proton, a significant fraction of the neutron's kinetic energy can be transferred in a single central collision. Materials containing hydrogen, e.g. polymers in various chemical compositions, are efficient neutron moderators and are often used as shielding materials. Regular concrete contains up to 10 \%$ _{weight} $ water molecules, and due to its higher mass density, it has comparable shielding efficiency for neutrons as polymers.

%

\section{Machine protection systems based on beam loss measurement}
\label{chapter-machine-protection}

\subsection{Requirements for machine protection systems}
\label{subsection-MPS-general}
The machine protection system (MPS) is a safety system installed at almost every accelerator; it creates a warning in case of a beam loss caused either by an irregular behaviour of a technical device or the~development of a beam instability \cite{Schmidt-CAS2014,Nordt-CAS2017,Wenn-CAS2016}.  An interlock is created to stop beam production in case of a LINAC or cyclotron, or initialize a beam dump by kicking the beam out of a synchrotron. Further beam production might be blocked until the cause of malfunction is eliminated.
There are three general requirements for the MPS:
\begin{itemize}
    \item \textit{Protection of the accelerator and environment:} Damage of the accelerator components must be avoided, which are either caused by the overheating (mostly due to the electronic stopping, see Sections~\ref{subsection-electronic-stopping} and \ref{subsection-material-heating}), radiation effects (due to material modifications) or activation (due to nuclear reactions, see Section~\ref{subsection-nuclear-stopping}). Moreover, the environment should be protected against unnecessary radiation exposure and activation. In case of doubt concerning correct operation, the MPS should generate an interlock to trigger a halt for beam operation.
    \item \textit{Protection of the beam:} The goal of any accelerator facility is to produce beams to serve users, which include, to an acceptable extent, the production of radiation and risk of damage. However, an interlock based on an incorrect device reading leads to an unnecessary stop of the beam and a~user's disturbance. Therefore, the MPS should be fail-safe and provide sufficient dynamic range for correct beam loss measurements.
    \item \textit{Providing the evidence:} In case of an interlock and related beam abortion, the cause must be assigned in a clear manner and archiving of relevant beam and device parameter must be provided by post-mortem archiving. With this information, a repetition of the same failure should be avoided.
\end{itemize}

An MPS delivers passive protection of components by measuring the fraction of the beam colliding with the vacuum pipe or limiting insertions. Two main categories for beam losses are distinguishable:
\begin{itemize}
    \item \textit{Irregular or fast losses} occur by a malfunction of an accelerator
    device, like a magnet power supplier or amplifier for an rf cavity, leading to a beam with wrong parameters, and, subsequently, the~loss of part or all beam particles. Moreover, a	misalignment of the device settings (e.g., non-centred beam inside a	quadrupole leading to steering) results in a beam mismatch at following accelerator stages, and part of the beam might be lost. The improper device setting can be initialized by incorrect automatic correction schemes or manual interventions by the operators. Moreover, as triggered by small beam variations, instabilities can lead to a noticeable beam loss, potentially destroying accelerator components. The task of MPS is to localize these losses, warning the operator or triggering an interlock to stop the beam delivery or initialize a beam abortion.
    \item \textit{Regular or slow losses} are accepted losses, e.g. at aperture limits of a collimator for halo cleaning or during injection and extraction from a synchrotron, and losses due to the finite lifetime of the~beam in a synchrotron. In many cases, the fraction of lost beam is only in the range of several \%. Most of the losses are unavoidable, but an increase in the loss rate is a hint of malfunctions.
\end{itemize}

Related to the irregular losses, the MPS must provide high sensitivity and an extended dynamic range. The high sensitivity is needed to provide a usable signal when only a tiny fraction of the beam is lost.  The high dynamic range is needed to detect irregular losses, which can easily vary by several orders of magnitude in signal strength, e.g. due to a power failure of a magnet or rf-cavity. The required dynamic range can be lower for typical regular losses due to the slower time variation of most processes. An example is the synchrotron injection or extraction optimization, which in many cases leads to beam losses, but this can be minimized. On the other hand, even small beam parameter changes might significantly increase beam loss at that limiting insertions; simulations of the loss mechanisms are required, see e.g. Ref. \cite{lech-PRAB2019}.

\subsection{Beam loss monitors and machine protection systems}
\label{subsection-MPS-hardware}

\begin{figure}
    \parbox{0.56\linewidth}{\centering
        \includegraphics*[width=88mm,angle=0]{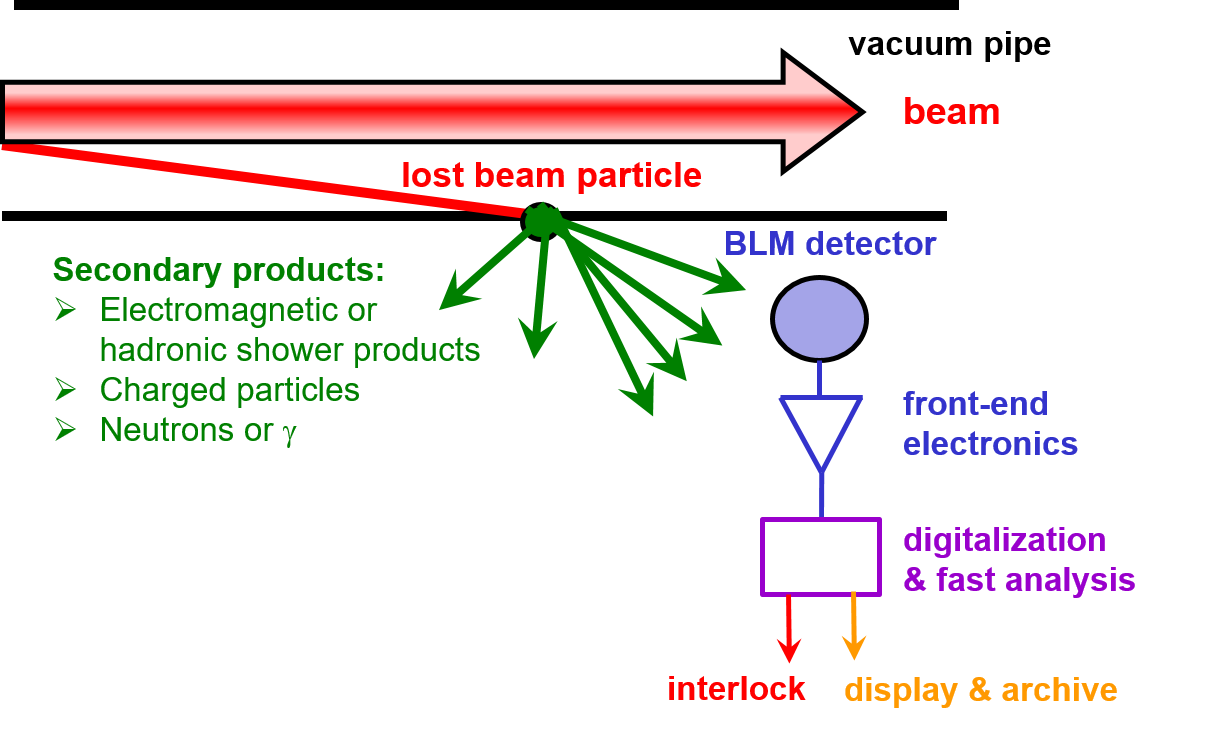} }
    \parbox{0.43\linewidth}{
        \includegraphics*[width=69mm,angle=0]{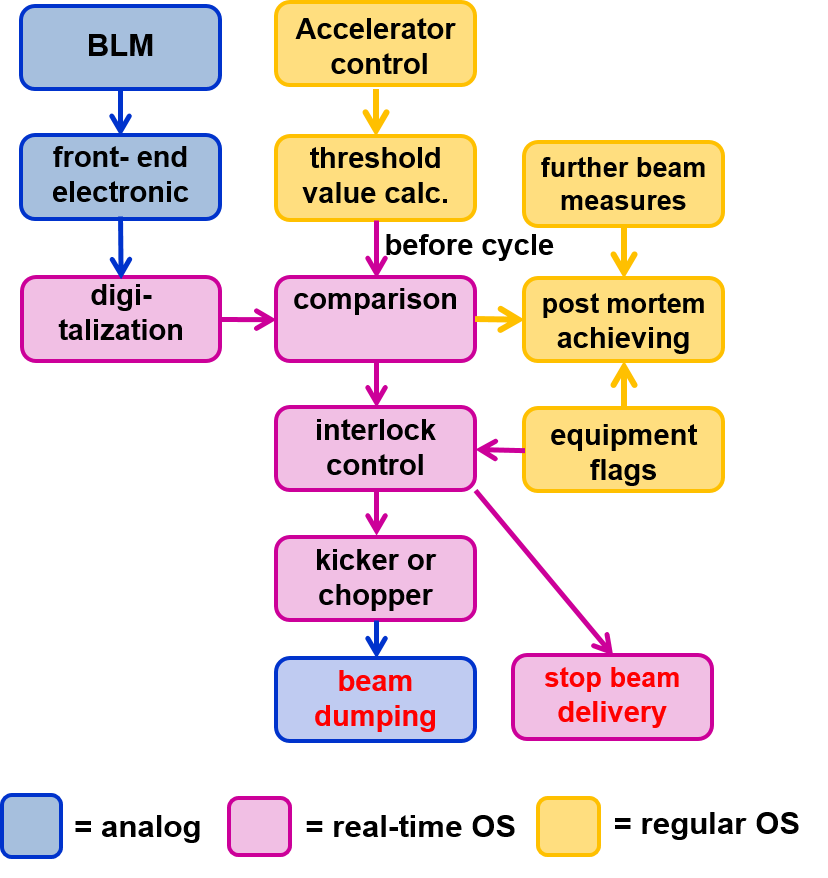} }
    \caption{General scheme of a machine protection system: Left: The beam loss monitor installation close to the~beam location. Right: Block diagram of the different components of the MPS, description see text.}
    \label{scheme-mps-all}
\end{figure}

A beam loss is characterized by the collision of the beam particle with the vacuum chamber or any other insertions as schematically depicted in Fig.~\ref{scheme-mps-all} (left). Depending on the beam properties, various reactions are possible as described in Sections~\ref{subsection-electron-stopping} to \ref{subsection-neutron-stopping}. In most cases, the beam particle enters the~vacuum chamber under grazing incident, leading to a large probability of interaction, i.e. the chamber acts as a thick target. Consequently, secondary particles are emitted; these can be charged particles like protons and electrons, neutrons, X-rays, or $\gamma$-rays. These secondary products are detected with beam loss monitors (BLM) installed outside of the vacuum chamber at appropriate locations with respect to the~expected shower maximum. The BLMs serve as a measurement device of the beam fraction scattering at this location. It requires a proper estimation of the beam particle interaction at the loss location, creating and transporting the secondary products, and their detection efficiency by the installed BLM type. Simulations are performed with codes such as FLUKA and GEANT4 \cite{fluka-www,geant4-www}.

Several types of BLMs are used as described in the frame of beam instrumentation either in this proceedings \cite{forck-CASgeneral} or more complete, e.g. in Refs. \cite{wittenburg_cas2018, dehning-CAS2016,dehning-ibic2012,zhuk-biw10,zhuk-pac13}. In brief, three types of BLMs are frequently used:
\begin{itemize}
    \item \textit{Ionization chamber:} An ionization chamber IC consists of a sealed Ar or N$_{2}$ gas volume with two or more electrodes biased by several kV. The electronic energy loss ionizes the gas during a charge particle passage, see Section~\ref{subsection-electron-stopping}. By biased electrodes, electrons and gas ions are separated, and the corresponding current is measured with sensitive electronics to provide the required dynamic range; examples are described in Refs. \cite{zhuk-biw10,dehning-ibic2012,stockner-dipac07}. The device measures directly the dose (energy deposition per mass of absorber) in the gas volume. It is mainly sensitive to charged particles.
    \item \textit{Scintillator:} Scintillators made of plastics are relatively cheap and available in many geometrical shapes. They can be operated in a single particle counting mode, which provides exceptionally high sensitivity and dynamic range. They are sensitive to charged particles and neutrons due to their elastic scattering at the polymers' hydrogen atoms, see Section~\ref{subsection-neutron-stopping}. Scintillators made of inorganic material are available as well, proving sensitivity to $\gamma$-rays. Scintillators are used if high sensitivity is needed, e.g. for regular loss determination at a collimator.
    \item \textit{Cherenkov counter:} At LINAC-based FEL facilities, a direct measurement of the lost electrons is of interest as it provides a low background measurement. Glass or inorganic crystals can be placed around the beam pipe; an example is given in Ref. \cite{yang-fel17}. As highly relativistic electrons' energy loss is low, they enter the glass with a velocity $v \simeq c$. Generally, Cherenkov radiation is emitted in case of a charge particle propagation with velocity $ v > c/n = c_{group}$ with $c$ is velocity of light and $n$ the index of refractivity of the medium \cite{leo-buch, jackson-buch, part-data-PhysRevD2018}.
\end{itemize}
The sensitivity, dynamic range and time response of the BLMs must be matched to the expected loss pattern, varying alongside the accelerator. Besides the detector principle, appropriate analogue electronics are used, and signals are digitized in such a manner to provide the required time resolution, sensitivity and dynamic range; an example for high-performance IC read-out described in Ref. \cite{venturini-ibic12}. Figure~\ref{scheme-mps-all} (right) depicts a general block diagram \cite{Schmidt-CAS2014,dehning-CAS2016,zhuk-biw10} of an entire MPS which comprises of:
\begin{itemize}
    \item \textit{Beam loss monitor front-end system:} The front-end system comprises the BLM, related analogue electronics and digitization; signal quality preservation is an important design criterion.
    \item \textit{Comparison to threshold losses:} The digitized values being proportional to the beam loss, is compared to the threshold of tolerance by a digital comparator. The reaction time for the interlock generation must fulfil the accelerator's time requirements, e.g. only a few $\mu$s for a proton LINAC. The threshold values are stored on this comparator within a look-up table to ensure fast access.
    \item \textit{Input from device discrepancy:} Besides the beam loss, any significant deviation of device parameter can trigger an interlock to stop the accelerator operation. For most devices (like power suppliers, rf-generators, vacuum measurements, pumps and valves), the actual values are monitored, and significant deviation from the set-values lead to a device interlock.
    \item \textit{Interlock control and ignition:} The sum-interlock is transferred to the abortion system controller handling the beam permit. The signal is passed to the power device, which chops the beam in case of a LINAC, prevent the beam delivery in case of a transport line, or trigger the emergency dump in case of a synchrotron. To ensure the required time response, all the above steps must be executed within a real-time hardware and software computer system. Real-time computing means that the response time is constant; part of the system is realized by hardware-related programming using logic blocks within field-programmable gate arrays (FPGA).
    \item \textit{Input of calculated maximal loss pattern:} As input to the comparator stage, threshold values for the actual operational modes is provided. As the thresholds might depend on the operation modes, e.g. for different beam energies and targets, they are calculated taking sophisticated models into account, see, e.g. Refs. \cite{Wenn-CAS2016,dehning-CAS2016,kallio-ipac15}. It can be executed beforehand on a regular, not real-time computer and stored in adequate databases. Before the beam execution, they are transmitted to the~look-up tables read by the real-time system.
    \item \textit{Display, data logging and archiving:} The reason for any beam abortion must be analyzed. The~required information at the time of the BLM threshold overshoot is stored together with other important beam parameter and device values to explain the incidence. The procedure is called a~post-mortem analysis.
\end{itemize}
A MPS is an extensive system as it contains many BLMs (e.g. more than 4000 at LHC), must handle several operation modes and related loss scenarios. It must be fail-safe on a very high level, to prevent for damages but does not lead to false interlocks. To check the functionality, hard- and software tests must be executed frequently, e.g. in short breaks between consecutive cycles. In case of doubts concerning the~functionality, the beam permit cannot be given. The redundancy of critical components might enlarge the reliability.

\subsection{Operation of a machine protection systems}
\label{subsection-MPS-operation}
The MPS priority is the protection of accelerator components related to heating, activation or material modifications by particles penetration as discussed in Section~\ref{section-beam-inter}. Even a tiny fraction of losses can lead to damage as the beam power at modern LINAC and synchrotron facilities are pretty high, see Fig.~\ref{beampower-all-acc}. For those beam parameters, the developments of a beam halo is expected \cite{witten-cas2018} as caused by non-ideal accelerator components, e.g. linear imperfections, third and higher-order field components of magnets, non-linearities of the acceleration rf, space-charge forces between the beam particles, and interaction with the residual gas within the vacuum pipe \cite{ziemann-cas-general, li-cas-general, tomas-cas2018, wenn-cas2018, cas-int-limit-2015}. Moreover, halos and beam losses can be caused by equipment uncertainties and beam instabilities. Equipment failures often lead to significant orbit executions, and consequently, a beam abortion is necessary. At synchrotron light sources, the emitted light can lead to overheating and damages.

Passive protection of sensitive components is realized by beam scrapers made of movable metal blocks driven closely to the beam path; they are also called collimator as several are installed within a beam-line or synchrotron. Particles within the beam halo are stopped in this metal if the scraper's longitudinal extension is longer than the particles' range. Movable scrapers longer than $\simeq$1~m are mechanically not feasible. Due to their longer range (see Fig.~\ref{copper_range}), protons above $E_{kin} > 1$~GeV are scattered out of the beam path and might be absorbed by the surroundings with the risk of activations. At LINAC facilities, the scraping is favourable at the lower energy part to minimize activations, e.g. discussed in Ref. \cite{plum-cas2014}. At synchrotrons, scrapers are installed at larger beam size locations, typically fulfilled at focusing quadrupoles for the horizontal plane and defocusing quadrupoles for the vertical plane. For LHC, such collimators are essential to protect the superconducting magnets and enable halo-free collisions at the user's detectors; the experiences are summarized in \cite{redaelli-CAS2016}. For the scrapers' position control, beam loss monitors record the shower particles related to the beam impact. In case of an irregular beam loss caused by device malfunction, an additional signal is recorded by these monitors and might trigger the~MPS to stop the beam production. For a correct interpretation in terms of beam loss fraction, numerical calculations by FLUKA, GEANT4 or comparable codes are required to estimate the secondary particles at the~BLM location; see, e.g. Ref. \cite{dehning-CAS2016}.

\begin{figure}
    \centering  \includegraphics*[width=110mm,angle=0]
    {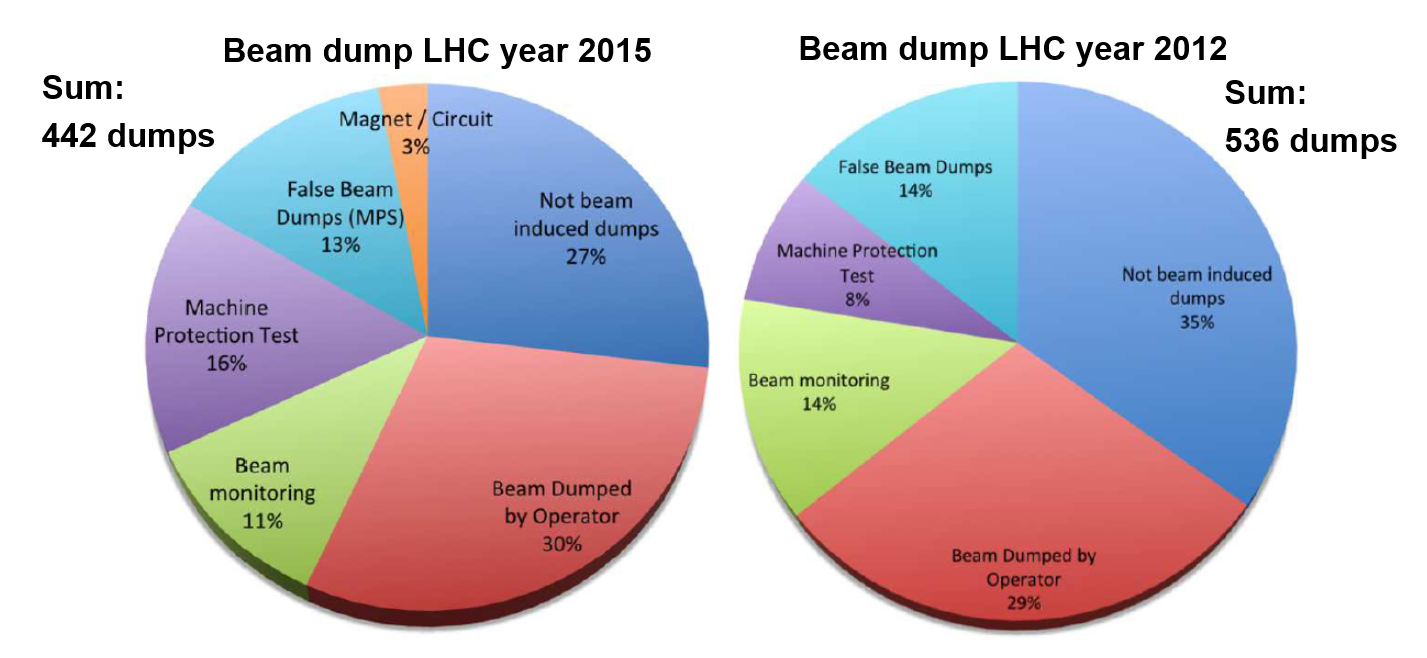}
    \caption{Statistics of beam abortions at LHC during the year 2015 (left) and 2012 (right), reproduced from \cite{woll-ipac2016}.}
    \label{LHC-dumps-wenninger}
\end{figure}

Figure~\ref{LHC-dumps-wenninger} shows an example the statistic of beam dumps at LHC for accelerated beams \cite{Wenn-CAS2016,todd-cern2013,woll-ipac2016,appol-ipac2018,zerlauth-ipac2019} documented by detailed post-mortem analysis. In the case of full functionality, the average storage time is $\simeq$10 h and dumped by the operators, while the mean storage time for unplanned abortions is $\simeq$5 h. The reasons for beam dumps are:
\begin{itemize}
    \item About 1/3 of the fills are terminated as planned, corresponding to regular LHC operation.
    \item About 1/3 of the dumps are caused by device (e.g. power suppliers, rf generators) failures.
    \item About 15 \% of the dumps are based on exceptional increased beam losses detected by the BLMs or orbit excursions, e.g. detected by Beam Position Monitors. These are the expected irregular losses that should cause an abortion.
    \item Tests of the MPS were executed, which enters in the statistics with 15 \% of the dumps.
    \item About 15 \% were unnecessary beam abortions caused false events as later on determined by post-mortem diagnosis.
\end{itemize}
Threshold values for the MPS are adapted to prevent magnet quenches (or other potential destructions) but allow for maximal machine availability by avoiding unnecessary dumps. Related to varying operational requirements, the dump statistics differ significantly over the years; see Ref. \cite{todd-cern2013}.

The commissioning phase of an accelerator always starts with a so-called pilot or witness beam of intensities below the damage threshold; this is realized by beams of lower current, short beam pulse at LINACs or the injection of only a few bunches in a synchrotron. With these beams, the device functionality is tested, and the MPS parameters are evaluated; an example is described in Ref. \cite{woll-ipac2016}. The exact MPS parameters must be balanced to be restrictive enough for damage prevention but allow flexible beam parameter changes during operation. The safety margin for the regular beam operation is evaluated during the commissioning period.

\newpage
\section{Protection of people}
\label{chapter-personal-protection}

\subsection{Radiological quantities and categories}
\label{subsection-radiological-quantities}

In this section the basic radiological quantities and thresholds for radiation areas are reviewed; more details are described on an introductory, but comprehensive level in Ref. \cite{grupen-book}. The basic physical quantity is the absorbed dose $D_{R,T}$ by the energy loss $\frac{\mbox{d}E_{R}}{\mbox{d}x}$ within a volume $V_{T}$ of mass $m_{T}$  and radiation type $R$ (electrons, ions, neutrons, $\alpha, \beta, \gamma$ etc.) as
\begin{equation}\label{dose}
    D_{R,T} = \dfrac{1}{m_{T}} \int_{V_{T}} \dfrac{\mbox{d}E_{R}}{\mbox{d}V}\;\mbox{d}V ~~~~~~\mbox{with the unit}~\left[\dfrac{\mbox{J}}{\mbox{kg}} \right] = \left[ \mbox{Gy} \right] = \left[ \mbox{100 rad} \right]
\end{equation}
in dependence of the tissue $T$. This dose is a measurable physical quantity with the main contribution related to charged particles' energy loss described in Section~\ref{subsection-electron-stopping}, and, corresponding sections for photons and neutrons.

The energy release of the radiation type $R$ is different, e.g. ions have high local energy deposition (see Fig.~\ref{copper_range}) and can destroy biological tissue more efficient than protons or electrons. Due to the strong interaction, neutrons scatter only on a nucleus, transferring some recoil momentum, which results in a large local energy deposition related to the electronic stopping of the recoil nucleus. Therefore, the damage of biological tissue is not expected to be proportional to the energy deposition; instead, the quantity equivalent dose $H_{T}$ is defined for each radiation type $R$ as
\begin{equation}\label{equi-dose}
    H_{T} =  \sum_{R} w_{R} D_{R,T} ~~~~~~\mbox{with the unit}~\left[ \mbox{Sv} \right] = \left[ \mbox{100 rem} \right]  ,
\end{equation}
Table~\ref{table-weight-R} compiles the radiation weighting factors $w_{R}$ for different organs. Based on biological investigations, the weighting factors are fixed by legal acts. Those values follow the recommendation of the~International Commission on Radiological Protection (ICRP) \cite{icrp} but might differ slightly due to adaptable national laws. For the conversion from physical to equivalent dose, the composition of the~radiation must be known, which is quite often not well determined at accelerators, as it depends on the~projectile-target combination and the applied shielding of the primary interaction point.

Human organs and tissues react quite different to irradiations. To describe the radiation damage, a~tissue weighting factor $w_{T}$ for each organ or tissue $T$ is introduced by the effective dose
\begin{equation}\label{effective-dose}
    E =  \sum_{T} w_{T} H_{T} ~~~~~~\mbox{with the unit}~\left[ \mbox{Sv} \right] = \left[ \mbox{100 rem} \right] .
\end{equation}
Examples of tissue weighting factors are compiled in Table~\ref{table-weight-T}; the corresponding organ value must be considered for partial human body irradiation. The normalization is $\sum_{T} w_{T} = 1$, corresponding to a~whole-body exposure. The $w_{T}$ factors show that irradiation of some parts of the human body should be avoided, which calls, e.g. for careful handling of radioactive sources. A large distance to a point-like source is a good precaution to avoid sensitive organs' exposure and minimize exposure time. Such personal behaviour follows the general ALARA principle, which stands for As Low As Reasonable Achievable. The values of tissue weighting factors follow ICRP recommendations but are fixed by national laws.

\begin{table}
    \parbox{0.45\linewidth}{
        \caption{Radiation weighting factors $w_{R}$\\according to German laws \cite{GermanBfS}.}
        \label{table-weight-R}
        \begin{tabular}{|l|r|}
            \hline \hline
            \textbf{Radiation type} ${\pmb R}$ &  ${\boldsymbol{w_{R}}}$\\ \hline
            X-ray and $\gamma$ all energies & 1\\
            Lepton: $e^{-}$, $e^{+}$, $\mu^{\pm}$ all energies & 1\\
            Proton:$E >2 $ MeV & 5\\
            $ \alpha$, heavy ions, all energies & 20\\
            Neutron: $E< 10$ keV & 5 \\
            $~~~~$ $10 \; \mbox{keV} < E < 100$ keV & 10 \\
            $~~~$ $100 \; \mbox{keV} < E < 2$ MeV & 20 \\
            $~~~~~$ $2 \; \mbox{MeV} < E < 20$ MeV & 10 \\
            $~~~~~~~~~~~~~~~~~~~~~$ $ E > 20$ MeV & 5 \\	\hline \hline
        \end{tabular}
    }
    \parbox{0.54\linewidth}{ \vspace*{-0.5cm}
        \caption{Sensitivity and tissue weighting factors $w_{T}$ for\\ several human organs according ICRP recommendations \cite{icrp}.}
        \label{table-weight-T}
        \begin{tabular}{|l|l|r|}
            \hline \hline
            \textbf{Organ or tissue} & \textbf{Sensitivity} & ${\boldsymbol{w_{T}}}$\\ \hline
            Gonads & very high & 0.20\\
            Lung, stomach, colon, marrow & high  & 0.12\\
            Hematopoietic system & high & 0.12\\
            Lymphatic system & high & 0.12\\
            Liver, esophagus, thyroid  & medium & 0.05\\
            Chest, other  & medium & 0.05\\
            Muscle, skin, bone surface  & low & 0.01\\
            whole-body exposure & medium & 1~~~~~~\\
            \hline \hline
    \end{tabular}
    }
\end{table}

\subsection{Measures for personal protection}
\label{subsection-areas-dosimeter}

To protect people against ionizing radiation, limits are defined: For the general public, the expected yearly equivalent dose rate by artificial installations must be below $H/t < 1$ mSv/a assuming a full year of exposure; this value corresponds to 0.5 $\mu$Sv/h for a duration of 2000 hours per year, i.e. for a typical working time. For radiation-exposed worker, like many employees at accelerator facilities, the threshold is organized in two categories: An equivalent dose rate of maximal 20 mSv/a, equals 10 $\mu$Sv/h for 2000 hours, is allowed for workers in category A assuming a whole-body exposure, while 6  mSv/a equals 3~$\mu$Sv/h for 2000 hours are allowed for workers in category B. For all employees, the maximal integrated life dose is 400Sv. Those numbers are subject to national acts and might differ between countries. For comparison: the estimated lethal dose is about 4000 mSv assuming a short time exposure. In Section \ref{subsection-natural-exposure} those numbers are compared to natural and artificial exposure.

At facilities with potential radiation, like accelerators, different access areas are defined \cite{grupen-book}:
\begin{itemize}
    \item \textit{Free access area}: Outside of the facilities with free public access, the equivalent dose above the~natural background must be below $H/t < 0.3$ mSv/a. The shielding of accelerators must be chosen to guarantee this threshold for all operational modes.
    \item \textit{Supervised area}: Within a supervised area, the radiation level must be measured permanently, and for person access, the equivalent dose must be below  $H/t < 3$ $\mu$Sv/h. Each person must wear a~passive dosimeter. In the case of exceedance, the access must be controlled.
    \item \textit{Controlled area}: In an area where radiation can occur above the limit of a supervised area, the access must be controlled, and each radiation-exposed worker must be equipped with a personalized, active (i.e. permanently readable) dosimeter. In the controlled area, the dose rate must be below $H/t < 25$ $\mu$Sv/h in any case.
    \item \textit{Limited access area}: For an area with potential dose rate up to $H/t < 100 $ $\mu$Sv/h, only limited access is possible, and personal registration is required; an active dosimeter is obligatory.
    \item \textit{Ruled access area}: For an area with a potential dose rate up to $H/t < 25 $ mSv/h, the access is limited for exceptional cases with a time limitation and protective equipment is strictly required.
    \item \textit{Prohibited access area}: If the potential dose rate is above $H/t > 25 $ mSv/h access is prohibited.
\end{itemize}
The thresholds follow ICRP recommendations but are fixed by national laws. The integrated dose over one month, one year and the life dose must not be exceeded. For children and pregnant women, significantly lower thresholds are relevant; see, e.g. Ref. \cite{grupen-book}.

Within the tunnel of most accelerators during beam operation, the radiation level is above the~threshold of a supervised area and is only accessible via gates, ensuring the stop of beam production. The~radiation is caused by lost primary particles and secondary ionizing radiation and activation, as described in Section \ref{section-beam-inter}. For low energy hadron accelerators providing energies below the Coulomb barrier, e.g. above 2.5 MeV for protons and above about 5 MeV/u for medium-heavy ions, no nuclear reactions are possible, and access to transfer lines might be allowed. A second important source of radiation is located at rf-cavities: Here, the strong electric field up to some MV/m might cause field emission of electrons from the surface and their acceleration. These electrons hit the opposite wall and create X-rays up to MeV energies that are not entirely shielded by the cavity walls. Therefore, a cavity can only be operated with full power if no person is in the controlled area. Activation from these X-rays is not expected, contrary to the beam-induced interaction.

The radiation level outside of the accelerator tunnel at locations of personal access is permanently measured, i.e. in the controlled area. The radiation outside of the concrete shield consists mainly of $\gamma$-rays and neutrons, as shown below in Section~\ref{subsection-shielding}. As a detector for $\gamma$-radiation in most cases a proportional tube is used \cite{knoll-buch, protection-part-data-PhysRevD2018,tho-handbook, grupen-book}, which provides a single particle $\gamma$-sensitivity for 30~keV~$<~E_{\gamma}~<$~1.5~MeV. For neutron detection so-called rem-counters \cite{knoll-buch, protection-part-data-PhysRevD2018,tho-handbook, grupen-book} are used: Neutrons within a first step are moderated by elastic collisions with hydrogen atom within polyethylene cylinder (see Section~\ref{subsection-neutron-stopping}) surrounding a proportional tube field with gas e.g. BF$_{3}$. Bor has a large cross-section $> 1$~kbarn neutron-induced fission via $^{10}_{~5}$B(n,$\alpha$)$^{7}_{3}$Li and a energy release (Q-value) of 2.3~MeV. The energy loss of the $\alpha$ and Li-nucleus trigger the tube discharge. Additionally, ionization chambers might be installed for physical dose normalization. In case of radiation exceedance, the beam is automatically stopped or dumped, and the next beam pulse is prohibited.

\begin{figure}
    \centering  \includegraphics*[width=155mm,angle=0]
    {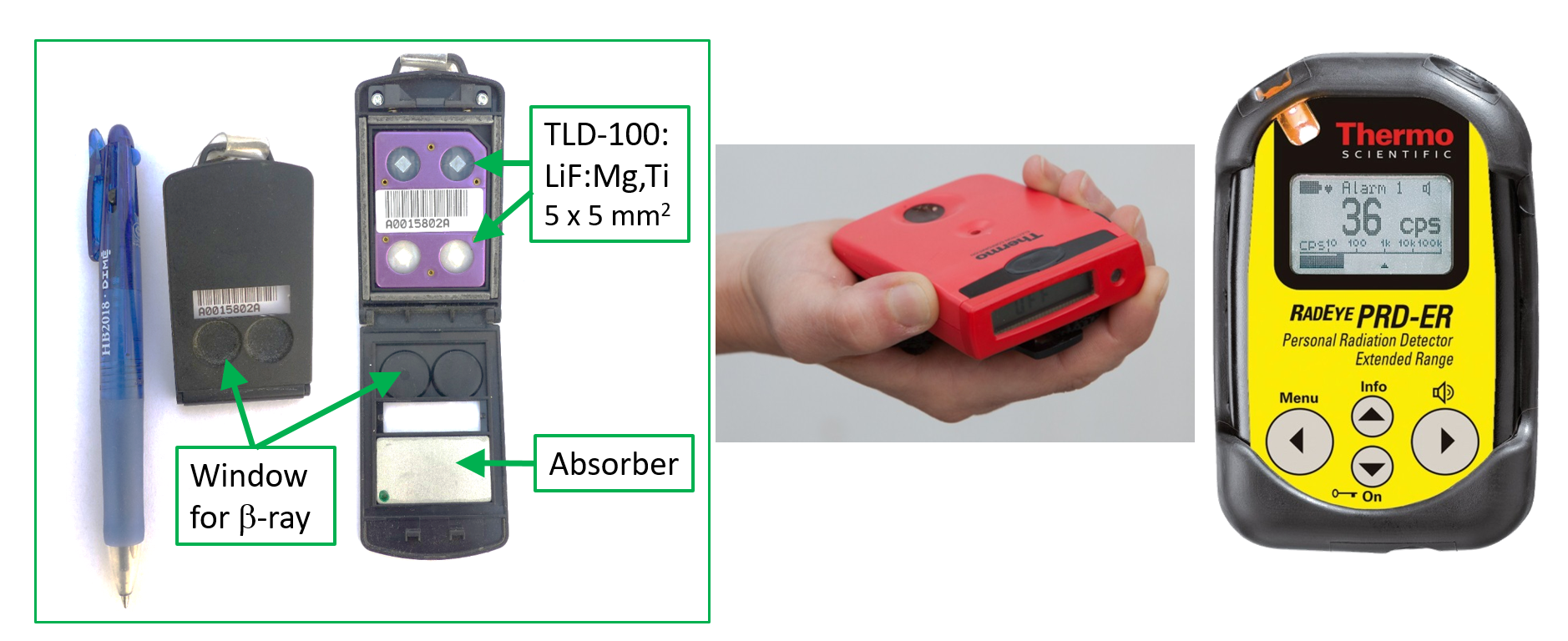} \vspace*{-3mm}
    \caption{Photos of several personal dosimeters: Left: pocket passive TLD dosimeter; middle: Geiger-M\"{u}ller tube based active dosimeter; right: NaI(Tl) scintillator based active dosimeter.}
    \label{photo-personal-dosimeters}
\end{figure}

A personal dosimeter must be worn to monitor the acquired dose for entrance to a controlled area and enable its personalized archiving. Passive dosimeters consist of thermoluminescent crystals, like LiF doped with Mg and Ti, so-called TLD-100 \cite{knoll-buch, grupen-book}. Thermoluminescence is the following process: Ionizing radiation leads to excitation of electrons from the valence-band to conduction-band of the LiF insulator; as influenced by the dopant, these electrons are trapped at atomic levels within the~band-gap. When the crystal is heated, typically to 200 $^{0}$C, these electrons are thermally excited and can recombine with valence-band holes by emission of optical light. This heating and light detection is executed monthly. Within a plastic case of typically 4 $\times$ 7 cm$^{2}$ several crystals of a typical size 5~$\times$ 5 mm$^{2}$  are placed and partly covered with absorbers to distinguish radiation types and monitor the~typical human skin dose, see Fig.~\ref{photo-personal-dosimeters}. Passive dosimeters are sensitive to $\beta$- and $\gamma$-rays; the equivalent dose sensitivity is about 100 $\mu$Sv integrated over one month. Significantly more sensitive are electrical handheld dosimeters, which are must be worn in limited access areas, see Fig.~\ref{photo-personal-dosimeters}. The dosimeter consists either of a Geiger-M\"{u}ller tube or NaI(Tl) scintillator \cite{knoll-buch, grupen-book}. They are sensitive to $\gamma$-rays with energies above 10 keV and $\beta$-radiation  above  300 keV as they must transfer the housing; $\alpha$ are not detected due to their short range. The dose rate threshold is typically 0.01~$\mu$Sv/h.

\subsection{Natural and artificial personal exposure}
\label{subsection-natural-exposure}
\begin{figure}
    \parbox{0.49\linewidth}{
        \centering  \includegraphics*[width=60mm,angle=0]
        {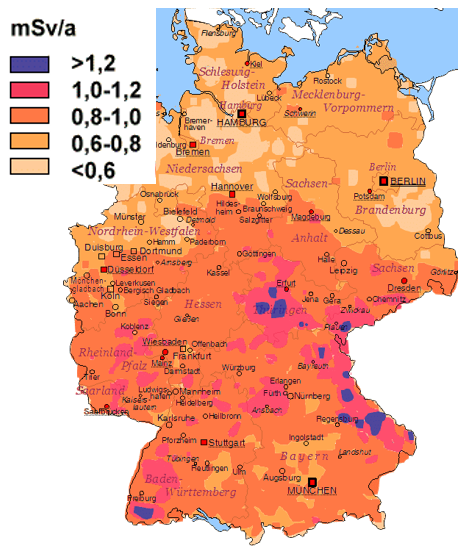}
    }
    \parbox{0.49\linewidth}{\centering
        \centering  \includegraphics*[width=60mm,angle=0]
        {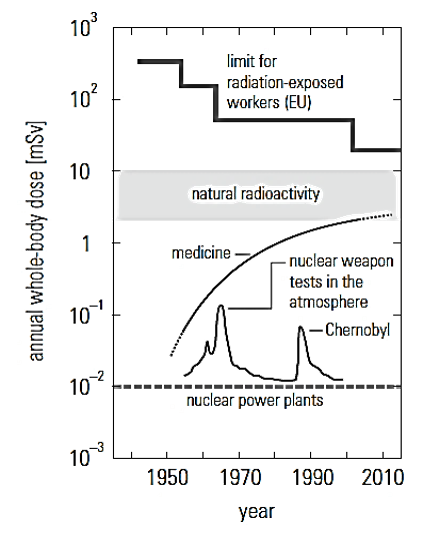}
    } \vspace*{-3mm}
    \caption{Left: Natural equivalent dose in Germany \cite{GermanBfS}. Right: Time evolution of a typical equivalent dose exposure in the past decades and the legal limit for radiation-exposed workers\cite{grupen-book}.}
    \label{people-exposure}
\end{figure}

Low radiation thresholds, defined in Section~\ref{subsection-areas-dosimeter}, are beneficial for radiation-exposed worker. To put it into perspective, the typical irradiation of a European inhabitant by natural and artificial sources is briefly summarized \cite{Forkel-CAS2013, grupen-book, cinelli-book}. Several sources originate natural exposure:
\begin{itemize}
    \item \textit{Cosmic radiation, $\simeq$0.3 mSv/a:} $\gamma$-rays and high energetic particles are generated in stars including our sun. Either they are reaching sea level directly, or, in most cases, interact within the earth atmosphere creating a shower of secondary particles \cite{cosmic-ray-part-data-PhysRevD2018}. The typical equivalent dose for a person in Europe is $\simeq0.3 $ mSv/a, with a lower value around the equator of $\simeq$0.2 mSv/a and higher at the pole region of $\simeq$0.5 mSv/a. The variation is originated from the guidance of charged cosmic particles by the earth magnetic field providing a higher field-line concentration towards the poles.
    \item \textit{Building material, $\simeq$0.5 mSv/a:} Several radiative isotopes (e.g. $^{40}_{19}$K, $^{226}_{~88}$Ra, $^{232}_{~90}$Th) are abundant in the earth surface, with large regional fluctuations related to geological conditions. As typical stony building materials produced from the local stone pits, contaminations by these isotopes lead to an average dose rate of $\simeq$0.5 mSv/a in dependence on the geographical region.
    \item \textit{Radon inhalation, $\simeq$1 mSv/a:} The rare gas $^{222}_{~86}$Rn is one product of the uranium and thorium decay chain taking place in the earth crust. As the rare gas radon is chemically inactive, it diffuses to the~surface and is emitted to the atmosphere. Related to the geological conditions, the concentration varies by more than one order of magnitude; in Fig.~\ref{people-exposure} (left), the regional fluctuations of radon irradiation in Germany is visible; for other countries, comparable maps exist \cite{cinelli-book}. The average dose for Europe is $\simeq$1 mSv/a.
    \item \textit{Incorporation, $\simeq$0.3 mSv/a:} Beside radon, other radioactive isotopes can be incorporated as they are present, e.g. in drinking water. A further example is $^{14}_{~6}$C, which is permanently produced by cosmic ray's secondary neutrons via the reaction $^{14}_{~7}$N(n,p)$^{14}_{~6}$C. This carbon isotope is incorporated in plants, animals and the human body. The total dose by inhalation and ingestion is on average $\simeq$0.3 mSv/a.
\end{itemize}
In summary, the natural equivalent dose for a European adult is about $\simeq$2 mSv/a with regional variations up to a maximum of $\simeq$10 mSv/a; the human body is adjusted to such irradiation.

In Fig.~\ref{people-exposure} (right), the average dose rate per year is shown as it developed in the past decades. Besides the constant natural exposure, a further contribution is related to artificial radioactivity by technical installations \cite{grupen-book,GermanBfS}:
\begin{itemize}
    \item \textit{Medical diagnostics, $\simeq$2 mSv/a}: X-ray diagnostics is the main contribution to the artificial exposure. Moreover, scintigrams for functional organ investigations are related to the large incorporation of radioactivity. In dependence on a person's health, the average equivalent dose for a~European is estimated to be $\simeq$2 mSv/a.
    \item \textit{Exposure during flights, up to $\simeq$10 $\mu$Sv/h}: Related to the shielding property of the~atmosphere, the~dose by cosmic rays in 10 km altitude is almost a factor 100 higher than on sea level. Caused by the earth magnetic field, the cosmic ray exposure depends on latitude and reaches at 11 km altitude up to $\simeq$10 $\mu$Sv/h; the sun activity with a $\simeq$10 year period leads to further fluctuations. A typical accumulated dose for a flight across the pole, e.g. from Europe to San Francisco, is $\simeq$45-110 $\mu$Sv, while a flight of similar duration across the equator, e.g. to Rio de Janeiro, the~accumulated dose is $\simeq$17-28 $\mu$Sv.
    \item \textit{Radiation by technical installations}: Under regular operational conditions, the release of radioactive isotopes from technical installation, like accelerators, nuclear or coal power plants, can be neglected. However, in case of failure, isotopes can be emitted in the atmosphere, as accidentally happened in the year 1986 in Chernobyl (Ukraine) and visible in Fig.\ref{people-exposure} (right). Moreover, atmosphere nuclear bomb tests contributed significantly; USA and Soviet Union tests were stopped in the year 1963 by a bilateral treaty; other countries continued with such tests.
\end{itemize}
The sum of natural and artificial equivalent dose exposure of a European person is $\simeq 3-5$ mSv/a; this should be compared to the 20 mSv/a threshold for radiation-exposed workers of category A defined in Section~\ref{subsection-areas-dosimeter}, which is regarded as a safe border for the maximal additional irradiation. At accelerator facilities, this border is only reached in very exceptional cases. However, any unnecessary exposure should be avoided by obeying the ALARA principle.

\subsection{Shielding of accelerators}
\label{subsection-shielding}
\begin{figure}
    \centering \includegraphics*[width=150mm,angle=0]
    {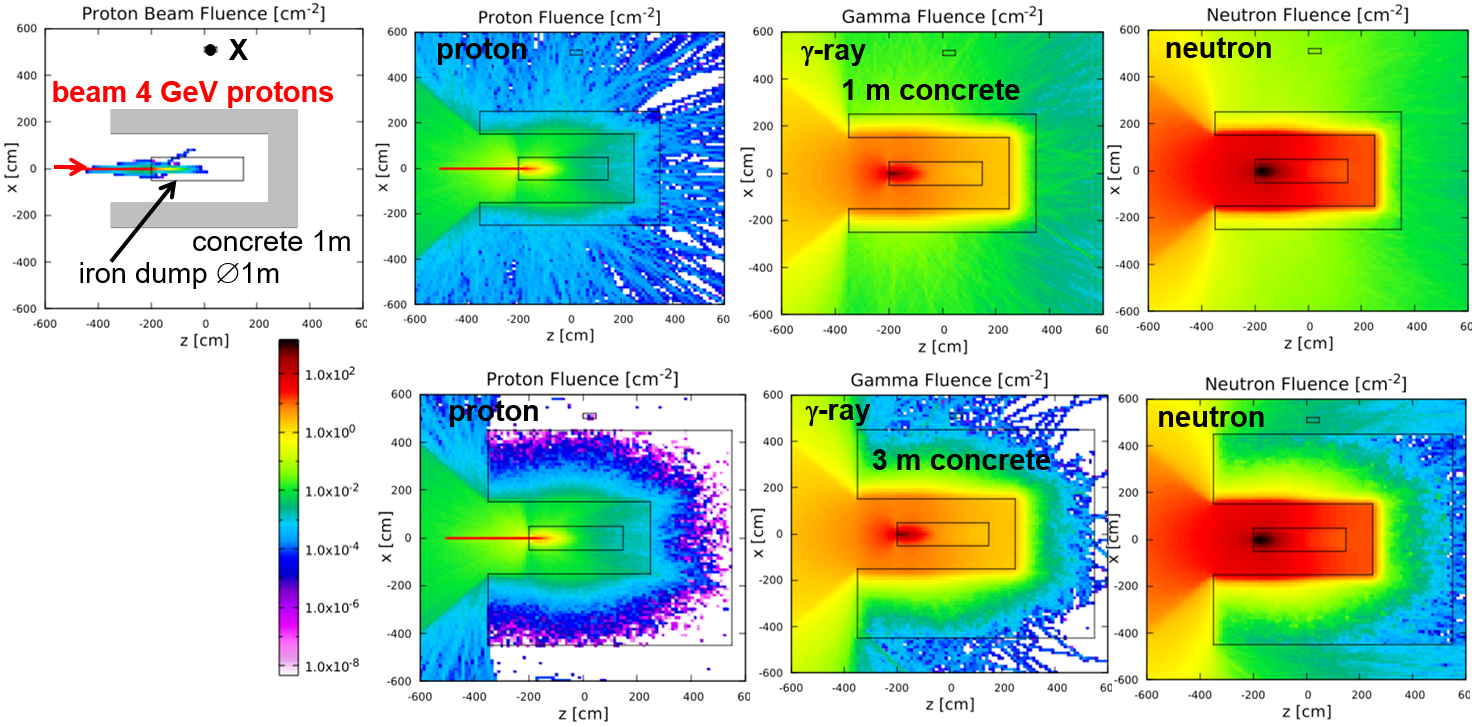}
    \caption{Simulation of a concrete shielding by FLUKA for a beam of $5 \times 10^{5}$ protons at 4 GeV stopped in a~3.5~m long iron beam dump \cite{udrea-priv}. The geometry is shown left; top row: 1 m, bottom row: 3 m u-shaped concrete shielding. The fluences of the assigned particles are logarithmically colour-coded.}
    \label{shielding-model-Fluka}
\end{figure}

Accelerated particles produce necessarily ionizing radiation when they are stopped in a target. To protect people, the accelerator area is shielded to absorb the secondary particles and to allow the designation of access areas (controlled, supervised or free access described in Section \ref{subsection-areas-dosimeter}). The physical basis related to the stopping of the beam, generation of secondary particles of all kind and their transport through matter is described in Section~\ref{section-beam-inter}. As these are extensive and multifaceted processes, numerical calculations are performed, in many cases using FLUKA \cite{fluka-www}. Moreover, the shielding concept is an integral part of the~governmental operation permit. The relevant physics, design rules and results are described in detail in Refs. \cite{Forkel-CAS2013, tho-handbook, sull-buch, shultis-buch, ncrp-report}; we discuss only one simplified example as depicted in Fig.~\ref{shielding-model-Fluka} to demonstrate some basic considerations.

The simplified model concerns the shielding properties of a beam dump, and the surrounding concrete of two different thicknesses as depicted in Fig.~\ref{shielding-model-Fluka} for the case of a 4 GeV proton beam. The~fluence in units [cm$^{-2}$] for the individual particles are depicted regardless of their energy. The beam impinges a 3.5 m long iron beam dump significantly longer than the protons' range (see Eq.~(\ref{eq-range-calc}) and Fig.~\ref{copper_range}) and a lateral extension of 1 m to ensure stopping of all primary protons. A cocktail of different particles is produced by spallation reactions as discussed in Section \ref{subsection-nuclear-stopping}. Moreover, those secondary particles might generate further free particles by their interaction in the dump and the walls. The qualitative behaviour for several particles are:
\begin{itemize}
    \item \textit{Protons:} Secondary protons are created by the spallation with dominance in froward direction but large transverse momentum compared to the primary beam. Most are efficiently decelerated within the iron beam dump by electronic stopping, see Eq.~(\ref{eq-bethe-bloch}). Due to the smaller lateral dump thickness, the emission distribution from the dump is peaked at some angle. The protons are further decelerated in concrete walls, and only a few sufficiently high energetic protons surpass the 1 m wall on almost straight trajectories, while basically all protons a shielded by the 3 m wall.
    \item \textit{Photons:} The $\gamma$-ray from the primary beam interaction are partly attenuated in the dump material made of medium-heavy element iron and leads to large-angle scattering as Compton effect dominates in the energy range $10~\mbox{keV} < E_{\gamma} < 10~\mbox{MeV}$, and manifests itself as the cone-shaped emission out of the surrounded area. The amount of $\gamma$s penetrating the concrete wall are attenuated by the 1 m thick wall by about two orders of magnitude compared to the fluence at the inner wall, but for the 3 m wall, the attenuation is larger, even though it would be more substantial using a~shielding made of heavier elements.
    \item \textit{Neutrons:} The spallation reaction creates a comparable number of protons and neutrons. Contrary to protons, elastic scattering dominates for neutrons, see Section~\ref{subsection-neutron-stopping}, resulting in small changes of energy but significant variations of angle. This results in an enhanced probability of backscattering, visible in Fig.~\ref{shielding-model-Fluka} by the cone leaving the shielded area. Moreover, in the concrete walls, reflections of the neutrons by elastic scattering on hydrogen nuclei have a high probability leading to a large density of neutrons within the area and contributes to the cone-shaped emission out of the shielded area. The attenuation of neutron number through the wall is more isotropic than for protons. As a rule-of-thumb, the transmission of neutrons through a concrete wall is roughly one order of magnitude per one meter of concrete.
\end{itemize}
As a result of these simplified considerations, it can be stated that protons have almost straight trajectories and are well shielded due to the electronic stopping. $\gamma$s are less attenuated and have large lateral straggling than protons; shielding made of heavy, high-$Z_{t}$ elements, e.g. metals, would increase the shielding. However, neutrons are not well shielded as elastic scattering is dominant and material of low weight is preferred; the elastic cross-section is large for scattering at hydrogen. At proton and ion accelerators, neutrons dominate radiation outside the thick shielding. For electrons accelerators, concrete walls are required as neutrons are produced in the dump and walls via Bremsstrahlung photons followed by giant resonances excitation as described in Section~\ref{subsection-photon-stopping}.

For a point source the attenuation of neutrons by different materials at a distance $d$ can roughly be approximated via the formula $H(d)= H_{0}/d^{2} \cdot 10 ^{-d/\lambda_{10}}$ with the magnitude attenuation constant $\lambda_{10} = \lambda_{\text{mfp}}  \cdot \ln 10$  related to the mean free path $\lambda_{\text{mfp}}$ of the neutrons; values for some frequently used shielding materials are compiled in Table~\ref{table-decrease-constant}. Regular concrete is often used due to the relatively low cost per volume. If possible, shielding by soil is used outside of buildings. Beam dumps are mostly made of iron as this offers reasonable neutron shielding and good efficiency for proton stopping related to the higher density with reasonable cost. Lead is a less efficient neutron shield but related to its high density it is frequently used for $\gamma$-rays and charged particles attenuation. Polymers can be alternative for neutron shielding. A shielding design based on numerical simulations includes the scattering of particles, the effect of activation, and the geometrical conditions.

\begin{table}
    \caption{Magnitude attenuation constant and density of some materials used for radiation shielding \cite{sull-buch,tho-handbook}.}
    \label{table-decrease-constant}
    \begin{center}
        \begin{tabular}{|l|r|r|r|r|r|} \hline \hline
            \textbf{Material} & \textbf{Soil} & \textbf{Concrete} & \textbf{Heavy Concrete} & \textbf{Iron} & \textbf{Lead}\\ \hline
            $\lambda_{10}$ [cm] & 128 & 100 & 80 & 41 & 39 \\
            density [g/cm$^{3}$]& 1.8 & 2.4 & 3.2 & 7.4 & 11.3\\ \hline \hline
        \end{tabular}
    \end{center}
\end{table}

\begin{figure}
    \centering   \hspace*{1cm} \includegraphics*[width=155mm,angle=0]
    {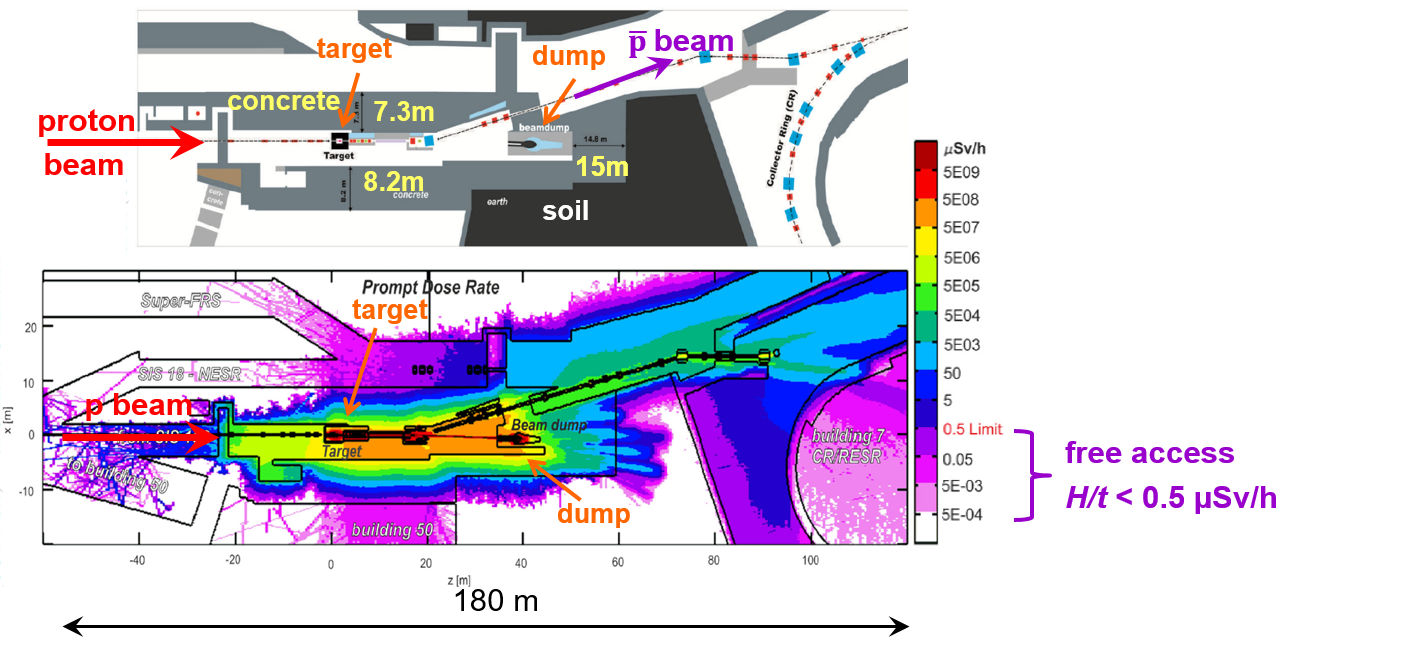} \vspace*{-7mm}
    \caption{Example of a FLUKA shielding simulation with a 11 cm long target irradiated by $2 \times 10^{13}$ protons of 29~GeV and a successive beam dump; the equivalent dose is logarithmically colour-coded \cite{knie-ipac12}.}
    \label{shielding-GSI-proton-target}
\end{figure}

As the shielding properties are part of an accelerator's operation permit, they are simulated in detail. One realistic example is depicted in Fig.~\ref{shielding-GSI-proton-target} for an anti-proton production target impinged by $2 \times 10^{13}$ proton of 29 GeV kinetic energy of $\simeq$100 ns duration \cite{knie-ipac12}: Within the nickel target, protons lose part of their kinetic energy and are stopped in a downstream beam dump. The target itself is surrounded by an iron block of 1.6 m thickness and a concrete housing. The associated transfer lines are shielded by a concrete wall of almost 10 m thickness.  In this example, the equivalent dose rate is calculated by extensive FLUKA \cite{fluka-www} simulations to model the interaction of primary and secondary particles in the surroundings. Much more than the lethal dose is reached in the 'hot zone' close to the target and beam dump. The thick concrete walls and additional soil are required for dose reduction. The dose is large behind the beam dump shielded by 15 m of concrete and soil. The radiation, mainly carried by neutrons, is still large within the area of transfer lines. Several labyrinths are visible for neutron reflections. The~shielding is designed for an average dose rate reduction to 0.5 $\mu$Sv/h for a free accessible area (purple and white coded), where other buildings are placed.

\section{Summary}
\label{chapter-conclusion}
Accelerators are built for basic and applied science, and ionizing radiation is produced on purpose. The~different physical aspects of the beam-matter interaction were discussed. One can distinguish between atomic physics interactions leading, as a first step, to free electrons and photons, following the charged particle's declaration. Secondary processes are the creations of high energetic photons, material heating and related damage, and modification of solid-state lattices, resulting in potential material dysfunction and restrictions of accelerator operation. Nuclear interactions lead to the creation of high energetic free particles of all kind. This includes neutrons that have long ranges in matter and require dedicated shielding considerations. Activation of accelerator components are caused by the nuclear interaction and hinder maintenance access. High power target, like for neutron, rare isotope, or anti-proton productions need remote control as the radiation level is above any safety threshold. Measurement methods and detectors for radiation detection were briefly introduced. Most important is to obey the ALARA principle, which can be applied for several topics:
\begin{itemize}
    \item Beam production shall be limited to the necessary amount for research and production.
    \item Collimators are often used to protect sensitive equipment and locate activation at dedicated areas.
    \item The risk of equipment damage shall be minimized by the machine protection system leading to controlled beam dumps.
    \item The environment shall be protected against activation and its disposal.
    \item The exposure of people must be strictly limited.
\end{itemize}
The production of radiation exposure must be weighted with the scientific, technical and eventually medical goals and achievements. A machine protection system's goal is to prevent dangerous situations by beam abortions followed by an analysis of the reasons. Legal rules for personal protection are stringent; exposure thresholds are extremely seldom reached at accelerator facilities.

\end{document}